\documentclass[prd,onecolumn,notitlepage,showpacs,preprintnumbers,amsmath,amssymb,nofootinbib,APS,11pt,superscriptaddress]{revtex4-1}

\usepackage{dcolumn}
\usepackage{bm}
\usepackage[applemac]{inputenc}
\usepackage[spanish,english]{babel}
\usepackage{amsmath,amssymb,amsfonts,latexsym,cancel}
\usepackage{graphicx}
\usepackage{color}
\usepackage{soul}
\usepackage{ulem}
\usepackage{hyperref}

\allowdisplaybreaks

\newcommand{\be}{\begin{equation}}
\newcommand{\ee}{\end{equation}}
\newcommand{\bea}{\begin{eqnarray}}
\newcommand{\eea}{\end{eqnarray}}

\begin{document}
\title{{\bf ADIABATIC REGULARIZATION WITH A YUKAWA INTERACTION}}

\author{Adrian del Rio}\email{adrian.rio@uv.es}
\author{Antonio Ferreiro}\email{antonio.ferreiro@ific.uv.es}
\author{Jose Navarro-Salas}\email{jnavarro@ific.uv.es}
\affiliation{Departamento de Fisica Teorica and IFIC, Centro Mixto Universidad de Valencia-CSIC. Facultad de Fisica, Universidad de Valencia, Burjassot-46100, Valencia, Spain.}
\author{Francisco Torrenti}\email{f.torrenti@csic.es}
\affiliation{Instituto de Fisica Teorica UAM/CSIC, Universidad Autonoma de Madrid, Cantoblanco, 28049 Madrid, Spain.}

\begin{abstract}

We extend the adiabatic regularization method for an expanding universe to  include the Yukawa interaction between quantized Dirac fermions and a homogeneous background scalar field.  We give explicit expressions for the renormalized expectation values of the stress-energy tensor $\langle T_{\mu\nu} \rangle$ and the bilinear $\langle \bar\psi\psi\rangle$ in a spatially flat Friedmann-Lemaitre-Robertson-Walker (FLRW) spacetime.  These are basic ingredients in the semiclassical field equations of  fermionic matter in curved spacetime interacting  with a background scalar field. The ultraviolet subtracting terms of the adiabatic regularization can be naturally interpreted as coming from appropriate counterterms of the background fields. We fix the required covariant counterterms. To test our approach we  determine the contribution of the Yukawa interaction to the conformal anomaly in the massless limit and show its consistency with  the heat-kernel  method using the  effective action. 

\end{abstract}


\date{\today}
\maketitle

\section{Introduction}\label{Introduction}

A major problem in the theory of quantized fields in curved spacetimes \cite{parker-toms, Waldbook, fulling, birrell-davies} is the computation of the expectation values of the stress-energy tensor components.  These calculations are rather convoluted, as they involve products of fields at coincident spacetime points, which are ultraviolet (UV) divergent even for free fields. In cosmological scenarios, this is connected to the fundamental phenomenon of particle creation by time-dependent backgrounds \cite{parker66, parker68}. A nonadiabatic expansion of the Universe generically induces particle creation of both bosonic and fermionic species, which leads to new UV divergences in their quadratic expectation values, not present in the absence of expansion.

A very efficient renormalization method, specifically constructed to deal with the UV divergences of a free field in an expanding universe, is adiabatic regularization.  Originally, this technique was introduced to tame the divergences of the mean particle number of a scalar field in a Friedmann-Lemaitre-Robertson-Walker (FLRW) universe \cite{parker66}, and was later extended to get rid of the divergences of the stress-energy tensor \cite{parker-fulling, Bunch80, Anderson-Parker}. The key ingredient of the adiabatic scheme is the asymptotic expansion of the field modes, in which increasingly higher-order terms in the expansion involve increasingly higher-order time derivatives of the metric (the scale factor). Due to dimensional reasons, this is equivalent to an UV asymptotic expansion in momenta. This way, one can expand adiabatically the integrand of the unrenormalized bilinear, identify the UV-divergent terms, and subtract them directly to obtain a  finite, covariant expre
 ssion. The renormalized expectation value is hence expressed as a finite integral in momentum space, depending exclusively on the mode functions defining the quantum state. The particular form of the adiabatic expansion depends on the spin of the quantized field. For free scalar fields, the well-known WKB expansion provides an adequate solution (see, for instance, \cite{parker-toms, fulling, birrell-davies}). For spin-1/2 fields, however, the adiabatic expansion takes a different form \cite{LNT,RNT} (see also \cite{Ghosh:2016}). The adiabatic method has been proven to be equivalent to the DeWitt-Schwinger point-splitting scheme \cite{christensen76, Christensen78} for both scalar fields \cite{Birrell, Anderson-Parker}, and spin-$1/2$ fields \cite{rio1}. The method is  specially suitable for numerical calculations, \cite{Hu-Parker, Anderson} and also for analytic approximations \cite{Anderson-Eaker, Molina-Paris-Anderson-Ramsey}.  The adiabatic regularization has also been use
 d to scrutinize the two-point function defining the variance and power spectrum in inflationary cosmology and related issues \cite{inflation-r, Woodard, Markkanen}.   
 We would like to note that, in cosmological perturbation theory, other fields apart from the inflaton are regarded themselves as first order. Then, in order to stay at linear order in cosmological fluctuations at the level of the equations of motion, it is necessary to study  these fields as propagating in the perfectly homogeneous FLRW background. This is what is customarily done in fact for the scalar curvature and tensor metric perturbations. Interactions between these metric perturbations and the additional fields are of higher order in cosmological perturbation theory. Therefore, when  adiabatic regularization (or any other renormalization method) is applied to inflationary cosmology, the additional fields  should be considered, at leading order, as quantum fields propagating in the homogeneous FLRW spacetime. 
%Both scalar  and tensor perturbations are treated this way. } 

Particle creation also takes place if the quantized field, either a boson or a fermion, is coupled to a classical background scalar field evolving nonadiabatically in time. In this case, the interaction term acts in the boson/fermion equation of motion as a time-dependent effective mass, which excites the field and increases its mean particle number. The most paradigmatic example of this is probably preheating after inflation \cite{KLS}. In the same way as before, new UV divergences associated to this particle creation appear in the expectation values of the different bilinears, not present in the absence of interaction, which must be appropriately removed to obtain  finite quantities. Although renormalization of expectation values for interacting fields is generally much more complicated, adiabatic regularization can be generalized to include interactions to classical scalar background fields. In this case, the adiabatic expansion of the field modes used to identify the UV-
 divergent terms depends on both the scale factor and the background field, as well as their respective time derivatives. If the quantized field is a scalar with a Yukawa-type coupling, the adiabatic expansion is still of the WKB form \cite{Anderson-Molina-Paris-Evanich-Cook,Molina-Paris-Anderson-Ramsey}. However, a generalization of the adiabatic scheme for other interacting species  in an expanding universe is absent  in the literature. Here, we will try to partially fill in this gap.

In this work, we extend the adiabatic regularization method to Dirac fields living in a FLRW universe and interacting, via the standard Yukawa coupling, with an external scalar field. In this approach, the Dirac field is quantized, while both the metric and the background field are regarded as classical. This kind of system appears for example in fermionic preheating, in which the inflaton acts as a background scalar field oscillating around the minimum of its potential, and decays nonperturbatively into fermions due to its Yukawa interactions  \cite{fermionic-preheating}. Another example is the decay of the Standard Model (SM) Higgs after inflation, in which the Higgs condensate oscillates around the minimum of its potential, and transfers part of its energy into all the massive fermions of the Standard Model, coupled to the Higgs with the usual SM Yukawa couplings \cite{figueroa,enqvist-meriniemi-nurmi} (another part  being transferred to the SM gauge bosons \cite{enqvist-
 meriniemi-nurmi,SMHiggsLattice}). In certain models, the Higgs decay may also lead to the reheating of the Universe \cite{higgs-preheating}. In this work, we will not focus on a particular scenario, but consider arbitrary time-dependent scale factors and background fields. The main objective is to provide well-motivated and rigorous expressions for the renormalized expectation values of the fermion stress-energy tensor $\langle T_{\mu\nu} \rangle$ and the bilinear $\langle \bar\psi\psi\rangle$. In the semiclassical equations of motion, these are the quantities that incorporate the backreaction of the created matter onto the background fields. To check the validity of the adiabatic method, we will also compute the contribution of the Yukawa interaction to the conformal anomaly in the massless limit, and check its consistency with the heat-kernel method using the effective action.

The paper is organized as follows. In Sec. \ref{semiclassicalequations} we give a general overview of the problem, introducing all necessary notation and  equations of motion of the system. In Sec. \ref{adiabaticexpansion} we develop the adiabatic expansion of the fermion field modes,  subject to a Yukawa interaction with a scalar background field. In Sec. \ref{renormalization} we derive general expressions for the renormalized expectation values  of the stress-energy tensor $\langle T_{\mu\nu} \rangle$ and the bilinear $\langle \bar\psi\psi\rangle$. In Sec. \ref{sectioncounterterms} we further analyze the  adiabatic regularization program by determining the covariant counterterms associated to the UV divergences. In Sec. \ref{sec:conformalanomaly} we apply the method to calculate the conformal anomaly, and include a discussion concerning the ambiguity of the coefficients of the anomaly on the renormalization scheme. We also show that our results are compatible
  with the heat-kernel method. Finally, in Sec. \ref{conclusions} we summarize our results and conclude. The paper is accompanied by three appendixes. In Appendix \ref{appendixA} we compute the conformal anomaly of a quantized scalar field coupled to a scalar background field. In Appendix \ref{appendixB} we apply adiabatic regularization to a simple, analytically solvable example. Finally, in Appendix \ref{appendixC} we  gather the terms of the fermionic adiabatic expansion up to fourth order. 

In this work, we take the FLRW metric as $ds^2= dt^2 -a^2(t) d\vec x^2$, and we use the Dirac-Pauli representation for the Dirac gamma matrices, $ 
\gamma^0 = \scriptsize
\left( {\begin{array}{cc}
 I & 0  \\
 0 & -I  \\
 \end{array} } \right)$, 
$\vec\gamma = \scriptsize \left( {\begin{array}{cc}
 0 & \vec\sigma  \\
 -\vec\sigma & 0  \\
 \end{array} } \right) $,
with $\vec{\sigma}$ the usual Pauli matrices. We also assume natural units $\hbar=1=c$.

\section{Semiclassical equations for a quantized Dirac matter  field  with Yukawa coupling }\label{semiclassicalequations}

We consider the theory defined by the action functional  $S=S[ g_{\mu\nu},\Phi,\psi,\nabla\psi]$,
where $\psi$ represents a  Dirac field, $\Phi$ is a scalar field, and $g_{\mu\nu}$ stands for the  spacetime metric. We decompose the action as $S=S_{g}+S_m$, where $S_m$ is the matter sector
\be \label{Sm} S_m = \int d^4x   \sqrt{-g} \  \left\{ \frac{i}{2}[ \bar \psi \underline\gamma^{\mu} \nabla_{\mu} \psi -(\nabla_{\mu} \bar \psi)\underline\gamma^{\mu} \psi)] - m\bar \psi \psi -g_Y \Phi \bar \psi \psi \right\} \ ,  \ee
and $S_g$ is the gravity-scalar sector, which will be presented in the next subsection. Here,  $\underline{\gamma}^{\mu}(x)$ are the spacetime-dependent Dirac matrices satisfying the anticommutation relations $\{\underline{\gamma}^{\mu},\underline{\gamma}^{\nu}\}=2g^{\mu\nu}$,  related to the usual Minkowski ones by the vierbein field $V_{\mu}^a(x)$ defined through $g_{\mu\nu}(x)=V_{\mu}^a(x)V_{\nu}^b(x) \eta_{ab}$. On the other hand, $\nabla_{\mu}\equiv \partial_{\mu}-\Gamma_{\mu}$ is the covariant derivative associated to the spin connection $\Gamma_{\mu}$, $m$ is the mass of the Dirac field, and $g_Y$ is the dimensionless coupling constant of the Yukawa interaction. In (\ref{Sm}), both the metric $g_{\mu\nu} (x)$ and the scalar  field $\Phi (x)$ are regarded as classical external fields. The Dirac spinor $\psi (x)$ will be our quantized field living in a curved spacetime and possessing  a Yukawa coupling to the classical  field $\Phi$. The Dirac equation is 
 \be (i  \underline\gamma^{\mu}\nabla_\mu -m - g_Y \Phi) \psi =0 \ , \ee
and the stress-energy tensor is given by \cite{birrell-davies}
\be T_{\mu\nu}^m :=\frac{2}{\sqrt{-g}}\frac{\delta S_m}{\delta g^{\mu\nu}}  =\frac{V_{\nu a}}{\det V}\frac{\delta S_m}{\delta V^{\mu}_{a}}   =  \frac{i}{2} \left[\bar \psi \underline\gamma_{(\mu}\nabla_{\nu)}\psi - (\nabla_{(\mu}\bar \psi)\underline\gamma_{\nu)} \psi \right]  \ .  \ee
The presence of the Yukawa interaction with the external field $\Phi$ modifies the standard conservation equation. We have, instead,
\be \label{covariance0}\nabla_\mu T_m^{\mu\nu}  = g_Y \bar \psi \psi  \nabla^{\nu} \Phi \ . \ee
These equations can be easily seen as the consequence of the invariance of the action functional $S$ under spacetime diffeomorphisms $\delta x^{\mu} = \epsilon^\mu(x)$: $\delta \Phi = \epsilon^{\mu}\nabla_{\mu} \Phi$, $\delta g_{\mu\nu}=2\nabla_{(\mu}\epsilon_{\nu)}$. 
One gets
\be \nabla_\mu T^{\mu\nu}_m + \frac{1}{\sqrt{-g}} \frac{\delta S_m}{\delta \Phi} \nabla^{\nu} \Phi =0 \ , \ee
which reproduces (\ref{covariance0}). We will assume that the quantum theory fully respects this symmetry. Therefore, we demand
\be \label{covariance}\nabla_\mu \langle T_m^{\mu\nu} \rangle = g_Y \langle \bar \psi \psi \rangle \nabla^{\nu} \Phi \ . \ee

\subsection{Adding the gravity-scalar sector}

The complete theory, including the gravity-scalar sector in the action, can be described by
\be S=  S_g + S_m = \frac{1}{16\pi G}\int d^4x\sqrt{-g} R + \int d^4x\sqrt{-g} \left\{ \frac{1}{2}g^{\mu\nu}\nabla_\mu \Phi \nabla_\nu \Phi
 - V(\Phi) \right\} + S_m \ ,  \label{eq:full-action}\ee
where $S_m$ is the action for the matter sector given in (\ref{Sm}). We will reconsider the form of the action in Sec. \ref{sectioncounterterms}, in view of the counterterms required to cancel the UV divergences of the quantized Dirac field. However, let us work for the moment with the action (\ref{eq:full-action}). The  Einstein equations are then
\be \label{eqg}G^{\mu\nu} +8\pi G (\nabla^\mu\Phi \nabla^\nu\Phi - \frac{1}{2}g^{\mu\nu} \nabla^\rho\Phi \nabla_\rho\Phi + g^{\mu\nu} V(\Phi)) = -8\pi G T_m^{\mu\nu} \ ,  \ee
and the equation for the scalar field is
\be \label{eqPhi}\Box \Phi + \frac{\partial V}{\partial \Phi} = - g_Y \bar \psi \psi \ . \ee
The semiclassical equations are obtained from (\ref{eqg}) and (\ref{eqPhi}) by replacing $ T_m^{\mu\nu}$ and $\bar \psi \psi$ by the corresponding (renormalized) vacuum expectation values $ \langle T_m^{\mu\nu}\rangle_{ren}$ and $\langle \bar \psi \psi \rangle_{ren}$,
\bea G^{\mu\nu} +8\pi G (\nabla^\mu\Phi \nabla^\nu\Phi - \frac{1}{2}g^{\mu\nu} \nabla^\rho\Phi \nabla_\rho\Phi + g^{\mu\nu} V(\Phi)) &=& -8\pi G \langle T_m^{\mu\nu} \rangle_{ren} \ , \label{eqg2b} \\*
\Box \Phi + \frac{\partial V}{\partial \Phi} &=& - g_Y \langle \bar \psi \psi \rangle_{ren}\ \label{eqPhi2}. \eea
These  equations  are consistent with the Bianchi identities  $\nabla_\mu G^{\mu\nu}=0$ , since
\be \nabla_{\mu}(\nabla^\mu\Phi \nabla^\nu\Phi - \frac{1}{2}g^{\mu\nu} \nabla^\rho\Phi \nabla_\rho\Phi + g^{\mu\nu} V(\Phi)) =  ( \Box \Phi + \frac{\partial V}{\partial \Phi})\nabla^\nu \Phi \ , \ee
and, from (\ref{covariance}) and (\ref{eqPhi2}), we have
\be \nabla_\mu \langle T_m^{\mu\nu} \rangle_{ren} = g_Y \langle \bar \psi \psi \rangle_{ren} \nabla^{\nu} \Phi = - ( \Box \Phi + \frac{\partial V}{\partial \Phi})\nabla^\nu \Phi \ . \ee

When the spacetime is an expanding universe [$ds^2= dt^2 -a^2(t) d\vec x^2$], and $\Phi$ is a homogeneous scalar field $\Phi = \Phi (t)$ (e.g. an inflaton), Eqs. (\ref{eqg2b}) and (\ref{eqPhi2}) describe the backreaction on the metric-inflaton system due to matter particle production and vacuum polarization codified in the renormalized  vacuum expectation values $\langle T_m^{\mu\nu} \rangle_{ren}$ and $\langle \bar \psi \psi \rangle_{ren}$. It is then important to elaborate an efficient method to compute these quantities  in this cosmological setting.

  \section{Adiabatic expansion for  a Dirac field with Yukawa coupling}\label{adiabaticexpansion}

In a spatially flat FLRW spacetime, the time-dependent gamma matrices are related with the Minkowskian ones by $\underline{\gamma}^0 (t) = \gamma^0$ and $\underline{\gamma}^i (t) = \gamma^i / a(t)$, and the components of the spin connections are $\Gamma_0=0$ and $\Gamma_i=(\dot{a}/2) \gamma_0 \gamma_i$.
The Dirac equation with the Yukawa interaction $i \underline{\gamma}^{\mu} \nabla_{\mu} \psi -  m \psi = g_Y \Phi \psi $, taking $\Phi$ as a homogenous scalar field $\Phi = \Phi(t)$, is then
\be \left( \partial_0 + \frac{3}{2} \frac{\dot a}{a} + \frac{1}{a} \gamma^0 \vec{\gamma} \vec{\nabla} + i (m+s(t)) \gamma^0 \right) \psi = 0 \ ,\label{direc}\ee
where we have defined $s(t) \equiv g_Y \Phi(t)$. If we expand the field $\psi$ as 
$ \psi=\int \frac{d^3 \vec{k}}{(2 \pi)^{3/2}} \psi_{\vec{k}}(t)e^{i\vec{k}\vec{x}}$,
and we substitute it into (\ref{direc}), we obtain the following differential equation for $\psi_{\vec{k}}$:
\be \left( \partial_t+ \frac{3\dot a}{2a} + i\gamma^0\vec{\gamma} \frac{\vec{k}}{a} +i\gamma^0 (m+s(t)) \right) \psi_{\vec{k}}=0 \label{psik-eq} \ .\ee
In order to solve this equation, it is convenient to write the Dirac field in terms of two two-component spinors of the generic form
\bea
&\psi_{\vec{k},\lambda}(t)= \frac{1}{a^{3/2} (t)}
\left( {\begin{array}{c}
  h^I_{{k}}(t) \xi_{\lambda} (\vec{k}) \\
  h^{II}_{{k}}(t)\frac{\vec{\sigma}\vec{k}}{k} \xi_{\lambda} (\vec{k})\\
 \end{array} } \right) \ ,
\eea
where  $\xi_{\lambda}$ with $\lambda ={\pm} 1$ are two constant orthonormal two-spinors ($\xi_{\lambda}^{\dagger}\xi_{\lambda'}=\delta_{\lambda,\lambda'}$), eigenvectors of the helicity operator $\frac{\vec{\sigma} \vec{k}}{2 k} \xi_{\lambda} = \frac{\lambda}{2} \xi_{\lambda}$. The explicit forms of  $\xi_{+1}$ and $\xi_{-1}$ are
\be \xi_{+1} (\vec{k}) = \frac{1}{\sqrt{2k (k + k_3)}}
\left( {\begin{array}{c}
 k + k_3  \\
 k_1 + i k_2  \\
 \end{array} } \right) \ ,
\hspace{1cm} 
\xi_{-1} (\vec{k}) = \frac{1}{\sqrt{2k (k + k_3)}}
\left( {\begin{array}{c}
 -k_1 + i k_2  \\
 k + k_3  \\
 \end{array} } \right) \ ,\ee
where $\vec{k}=(k_1, k_2, k_3)$ and $|\vec{k}| = k$. 
The time-dependent functions $h_k^{I}$ and $h_k^{II}$ satisfy the first-order coupled equations
\be  h_k^{II} = \frac{i a }{k} \left( \frac{\partial h_k^I}{\partial t} + i (m + s(t))h_k^I \right) \ , \,\,\,\,\,\,\,\,\,\, h_k^{I} =  \frac{i a }{k} \left( \frac{\partial h_k^{II}}{\partial t} - i (m + s(t))h_k^{II} \right) \label{ferm-hk2b} \ . \ee
Given a particular solution $\{ h_k^{I}(t)$, $h_k^{II}(t) \}$ to Eqs. (\ref{ferm-hk2b}),  one can construct the modes
\bea \label{uk}
&u_{\vec{k},\lambda}(t)= \frac{e^{i\vec k \vec x}}{\sqrt{(2\pi)^3a^{3} (t)}}
\left( {\begin{array}{c}
  h^I_{{k}}(t) \xi_{\lambda} (\vec{k}) \\
  h^{II}_{{k}}(t)\frac{\vec{\sigma}\vec{k}}{k} \xi_{\lambda} (\vec{k})\\
 \end{array} } \right) \ .
\eea
Equation (\ref{uk}) will be a solution of positive-frequency type in the adiabatic regime. A solution of negative-frequency type can be obtained by applying a charge conjugate transformation $C\psi = -i\gamma^2\psi^*$ (we follow here the convention in \cite{Peskin-Schroeder})
\bea \label{vk}
&v_{\vec{k},\lambda}(t)=C u_{\vec{k},\lambda}(t)=\frac{e^{-i\vec k \vec x}}{\sqrt{(2\pi)^3a^{3} (t)}}
\left( {\begin{array}{c}
  h^{II*}_{{k}}(t) \xi_{-\lambda} (\vec{k}) \\
  h^{I*}_{{k}}(t)\frac{\vec{\sigma}\vec{k}}{k} \xi_{-\lambda} (\vec{k})\\
 \end{array} } \right) \ .
\eea
The Dirac inner product is defined as $ (\psi_1,\psi_2)=\int d^3x a^3 \psi_1^{\dagger} \psi_2$. The normalization condition for the above four-spinors,  $(u_{\vec{k}\lambda},v_{\vec{k}\,'\lambda'})=0$,  $(u_{\vec{k}\lambda},u_{\vec{k}\,'\lambda'})=(v_{\vec{k}\lambda},v_{\vec{k}\,'\lambda'})=\delta_{\lambda\lambda'}\delta^{(3)}(\vec{k}-\vec{k}\,') $, reduces to 
\be |h_k^{I}|^2 + |h_k^{II}|^2=1 \ . \label{ferm-wronsk2} \ee 
Since the  Dirac scalar product  is preserved by the cosmological evolution, the normalization condition (\ref{ferm-wronsk2}) holds at any time. This ensures also the standard anticommutation relations for the creation and annihilation operators [$\{ B_{\vec{k},\lambda} , B_{\vec{k'},\lambda'}^{\dagger} \} = \delta^{3} (\vec{k} - \vec{k'} ) \delta_{\lambda \lambda'} $, $\{ B_{\vec{k},\lambda} , B_{\vec{k'},\lambda'}\} = 0$, and similarly for the  $D_{\vec{k},\lambda}$, $D_{\vec{k'},\lambda'}^{\dagger}$ operators], defined by the Fourier expansion of the Dirac field operator 
\bea
\psi(x)=\int d^3\vec{k} \sum_{\lambda}\left[B_{\vec{k} \lambda}u_{\vec{k} \lambda}(x)+D_{\vec{k} \lambda}^{\dagger}v_{\vec{k} \lambda}(x) \right] \ .  \label{4c}
\eea

\subsection{Adiabatic expansion}

We now compute the adiabatic expansion of a Dirac field living in a FLRW spacetime, and possessing a Yukawa interaction term with a classical background field. We know that, in the adiabatic limit, and in the absence of interaction, the natural solution of the field modes $h_k^I$ and $h_k^{II}$ is
 \be h_k^{I} (t) = \sqrt{\frac{\omega(t) + m}{2 \omega(t)}} e^{-i \int^t \omega(t') dt'}\ , \,\,\,\,\,\,\,\,\,\,\,\,\,\,\,\, h_k^{II} (t) = \sqrt{\frac{\omega(t) - m}{2 \omega(t)}} e^{-i \int^t \omega(t') dt'} \ , \label{ferm-minkowski} \ee
 where $\omega=\sqrt{\frac{k^2}{a^2}+m^2}$ is the frequency of the field mode. This will constitute the zeroth-order term of the adiabatic expansion. Mimicking the ansatz introduced in \cite{LNT}, we write the $h_k^I$ and $h_k^{II}$ functions as
  \be h_k^{I} (t) = \sqrt{\frac{\omega(t) + m}{2 \omega(t)}} e^{-i \int^t \Omega(t') dt'} F(t) \ , \,\,\,\,\,\,\,\,\,\,\,\,\,\,\,\, h_k^{II} (t) = \sqrt{\frac{\omega(t) - m}{2 \omega(t)   }} e^{-i \int^t \Omega(t')  dt'} G(t) \ , \label{ferm-ansatz} \ee
where $\Omega(t)$, $F(t)$, and $G(t)$ are time-dependent functions, which we expand adiabatically as
\bea \Omega &=& \omega + \omega^{(1)} + \omega^{(2)} + \omega^{(3)} + \omega^{(4)} + \dots \ , \nonumber \\*
F &=& 1 + F^{(1)} + F^{(2)} + F^{(3)} + F^{(4)} + \dots \ ,\nonumber \\*
G &=& 1 + G^{(1)} + G^{(2)} + G^{(3)} + G^{(4)} + \dots \ . 
\label{ferm-expansions} \eea
Here, $F^{(n)}$, $G^{(n)}$, and $\omega^{(n)}$ are terms of $n$th adiabatic order (we explain exactly what we mean by that below). By substituting (\ref{ferm-ansatz}) into the equations of motion (\ref{ferm-hk2b}) and the normalization condition (\ref{ferm-wronsk2}), we obtain the following system of three equations,
\bea 
(\omega - m) G = \Omega F + i \dot{F} + \frac{i F}{2} \frac{d \omega}{dt} \left( \frac{1}{\omega + m} - \frac{1}{\omega} \right) - (m + s) F \ , \nonumber \\*
(\omega + m) F = \Omega G + i \dot{G} + \frac{i G}{2} \frac{d{\omega}}{dt} \left(\frac{1}{\omega - m} - \frac{1}{\omega} \right) + (m + s) G \ , \nonumber \\*
(\omega + m) F F^{*} + (\omega - m) G G^{*} = 2 \omega \ . \label{system3q}
\eea
To obtain the expressions for $\Omega^{(n)}$, $F^{(n)}$, and $G^{(n)}$, we introduce the adiabatic expansions (\ref{ferm-expansions}) into (\ref{system3q}), and solve order by order. As usual, we consider $\dot a$ of adiabatic order one, $\ddot a$ of adiabatic order two, and so on. On the other hand, we consider the interaction term $s(t)$ of adiabatic order one, so that the zeroth-order term  in (\ref{ferm-ansatz})  recovers the free field solution in the adiabatic limit, defined in (\ref{ferm-minkowski}). Similarly, time derivatives of the interaction increase the adiabatic order, so that $\dot s$ is of order two, $\ddot s$ of order three, and so on. With this, a generic expression $f^{(n)}$ of adiabatic order $n$ (e.g. $f^{(n)}=F^{(n)},G^{(n)},\Omega^{(n)}$) will be written as a sum of all possible products of $n$th adiabatic order formed by $s$, $a$, and their time derivatives. For example, functions of adiabatic orders one and two will be written respectively as
\bea f^{(1)} &=& \alpha_1 s + \alpha_2 \dot a \ , \nonumber \\*
f^{(2)} &=& \beta_1 s^2 + \beta_2 \dot s + \beta_3 \ddot a + \beta_4 \dot a^2 + \beta_5 \dot a s \ , \eea
with $\alpha_n \equiv \alpha_n (m, k, a)$ and $\beta_n \equiv \beta_n (m, k, a)$. The assignment of $s$ as adiabatic order one is consistent with the scaling dimension of the scalar field, as it possesses the same dimensions as $\dot a$. 

\subsubsection{First adiabatic order}

By keeping only terms of first adiabatic order in (\ref{system3q}), the system of three equations gives
\bea
(\omega - m) G^{(1)} = (\omega - m) F^{(1)} + \omega^{(1)} - s + \frac{i}{2} \frac{d \omega}{dt} \left( \frac{1}{\omega + m} - \frac{1}{\omega} \right) \nonumber \ , \\*
(\omega + m) F^{(1)} = (\omega + m) G^{(1)} + \omega^{(1)} + s + \frac{i}{2} \frac{d \omega}{d t} \left( \frac{1}{\omega - m} - \frac{1}{\omega} \right) \nonumber \ , \\*
(\omega + m) (F^{(1)} + F^{(1) *} ) + (\omega - m ) (G^{(1)} + G^{(1) *} ) = 0 \ .
\eea
We now treat independently the real and imaginary parts by writing $F^{(1)} = f_x^{(1)} + i f_y^{(1)}$ and $G^{(1)} = g_x^{(1)} + i g_y^{(1)}$. We obtain for the real part
\bea
(\omega - m) (g_x^{(1)} - f_x^{(1)} ) &=& \omega^{(1)} - s \ , \nonumber \\*
(\omega + m) (g_x^{(1)} - f_x^{(1)} ) &=& - \omega^{(1)} - s  \ , \nonumber \\*
(\omega + m) f_x^{(1)} + (\omega - m) g_x^{(1)} &=& 0 \ ,
\eea
which has as solutions
\be f^{(1)}_x = \frac{s}{2 \omega} - \frac{m s}{2 \omega^2} \ , \,\,\,\,\,\,\,\,\,\,\,\, g^{(1)}_x = - \frac{s}{2 \omega} - \frac{m s}{2 \omega^2} \ , \,\,\,\,\,\,\,\,\,\,\,\, \omega^{(1)} = \frac{m s}{\omega} \ .\ee
On the other hand, the imaginary part of the system gives
\bea
(\omega - m) (g_y^{(1)} - f_y^{(1)} ) &=& \frac{1}{2} \frac{d \omega}{dt} \left( \frac{1}{\omega + m} - \frac{1}{\omega} \right) \ , \nonumber \\*
(\omega + m) (g_y^{(1)} - f_y^{(1)} ) &=& - \frac{1}{2} \frac{d \omega}{dt} \left( \frac{1}{\omega - m} - \frac{1}{\omega} \right) \ .
\eea
These two equations are not independent. The obtained solution for $g_y^{(1)}$ and $f_y^{(1)}$ is
\be f^{(1)}_y = A - \frac{m \dot a}{2 a \omega^2} \ , \,\,\,\,\,\,\,\,\,\,\,\, g^{(1)}_y = A \ , \ee
where $A$ is an arbitrary first-order adiabatic function. 
We will choose the simplest solution 
\be f^{(1)}_y = - \frac{m \dot{a}}{4 \omega^2 a} \ , \,\,\,\,\,\,\,\,\,\,\,\,  g^{(1)}_y = \frac{m \dot{a}}{4 \omega^2 a} \ ,\ee
obeying the condition
 $F^{(1)} (m, s) = G^{(1)} (-m, -s)$. Therefore, the adiabatic expansion will also preserve the symmetries of Eqs. (\ref{ferm-hk2b}) with respect to the change $(m,s) \rightarrow (-m,-s)$. We have checked that physical expectation values are independent to any potential ambiguity in this kind of  choice.

\subsubsection{Second adiabatic order}

In the same way, the second-order terms of (\ref{system3q}) give
\bea 
(\omega - m) G^{(2)} = (\omega - m) F^{(2)} + (\omega^{(1)} - s) F^{(1)} + \omega^{(2)} + i \dot{F}^{(1)} + i \frac{F^{(1)}}{2} \frac{d \omega}{dt} \left( \frac{1}{\omega + m} - \frac{1}{\omega} \right) \ , \nonumber \\*
(\omega + m) F^{(2)} = (\omega + m) G^{(2)} + (\omega^{(1)} + s) G^{(1)} + \omega^{(2)} + i \dot{G}^{(1)} + i \frac{G^{(1)}}{2} \frac{d \omega}{dt} \left( \frac{1}{\omega - m} - \frac{1}{\omega} \right) \ , \nonumber \\*
(\omega + m) (F^{(2)} + F^{(1)} F^{(1)*} + F^{(2)* }) + (\omega - m) (G^{(2)} + G^{(1)} G^{(1)*} + G^{(2)*} ) = 0 \ ,
\eea
where the first-order terms have already been  deduced above. Taking the real part of these equations, we obtain
\bea
(\omega - m) (g_x^{(2)} - f_x^{(2)} ) = (\omega^{(1)} - s) f_x^{(1)} + \omega^{(2)} - \dot{f}_y^{(1)} - \frac{f_y^{(1)}}{2} \frac{d \omega}{dt} \left( \frac{1}{\omega + m} - \frac{1}{\omega} \right) \ , \nonumber \\
(\omega + m) (g_x^{(2)} - f_x^{(2)} ) = - (\omega^{(1)} + s) g_x^{(1)} - \omega^{(2)} + \dot{g}_y^{(1)} + \frac{g_y^{(1)}}{2} \frac{d \omega}{dt} \left( \frac{1}{\omega - m} - \frac{1}{\omega} \right) \ , \nonumber \\
(\omega + m) (2 f_x^{(2)} + (f_x^{(1)})^2 + (f_y^{(1)})^2 ) + (\omega - m) (2 g_x^{(2)} + (g_x^{(1)})^2 + (g_y^{(1)})^2 ) = 0 \ ,
\eea
which has as solutions
\bea f^{(2)}_x &=& \frac{m^2 \ddot{a}}{8 a \omega^4}-\frac{m \ddot{a}}{8 a \omega^3}-\frac{5 m^4 \dot{a}^2}{16 a^2 \omega^6}+\frac{5 m^3 \dot{a}^2}{16 a^2 \omega^5}+\frac{3 m^2 \dot{a}^2}{32 a^2 \omega^4}-\frac{m \dot{a}^2}{8 a^2 \omega^3}+\frac{5 m^2 s^2}{8 \omega^4}-\frac{m s^2}{2 \omega^3}-\frac{s^2}{8 \omega^2} \ , \nonumber \\*
\omega^{(2)} &=& \frac{-m^2 s^2}{2 \omega^3} + \frac{s^2}{2 \omega} + \frac{5 m^4 \dot{a}^2}{8 a^2 \omega^5} - \frac{3 m^2 \dot{a}^2}{8 a^2 \omega^3} - \frac{m^2 \ddot{a}}{4 a \omega^3} \ ,\eea
and $g^{(2)}_x (m, s) = f^{(2)}_x (-m, -s)$. On the other hand, taking the imaginary part of the equations, we have
\bea
(\omega - m) (g_y^{(2)} - f_y^{(2)} ) &=& (\omega^{(1)} - s ) f_y^{(1)} + \dot{f}_x^{(1)} + \frac{f_x^{(1)}}{2} \frac{d \omega }{ dt} \left( \frac{1}{\omega + m} - \frac{1}{\omega} \right)  \ , \nonumber \\*
(\omega + m) (g_y^{(2)} - f_y^{(2)} ) &=& - (\omega^{(1)} + s ) g_y^{(1)} - \dot{g}_x^{(1)} - \frac{g_x^{(1)}}{2} \frac{d \omega }{ dt} \left( \frac{1}{\omega - m} - \frac{1}{\omega} \right)  \ .
\eea
As before, this system contains an arbitrariness in its solution,
\be f_y^{(2)} = B + \frac{5 m^2 s \dot{a}}{4 a \omega^4} - \frac{s \dot{a}}{2 a \omega^2} - \frac{\dot{s}}{2 \omega^2} \ , \hspace{0.5cm} g_y^{(2)} = B  \ , \ee
where now $B$ is a linear combination of second-order adiabatic terms. By imposing again the condition $F^{(2)} (m, s) = G^{(2)} (-m, -s)$, one finds
\be f^{(2)}_y = \frac{5 m^2 s \dot{a}}{8 a \omega^4} - \frac{s \dot{a}}{4 a \omega^2} - \frac{\dot{s}}{4 \omega^2} \ , \ee
and $g_y^{(2)} (m, s) = f_y^{(2)} (-m, -s)$. 

\subsubsection{Third and fourth adiabatic order}

The same procedure can be repeated for all orders. The real part of the expansion is totally determined by the system of equations (\ref{system3q}), while every imaginary part contains an arbitrariness that can be solved by fixing the condition $F^{(n)} (m, s) = G^{(n)} (-m, -s)$. The third- and fourth-order terms of the expansion are explicitly written in Appendix \ref{appendixC}.

\section{Renormalization of  the stress-energy tensor $\langle T_{\mu\nu} \rangle$ and the bilinear $\langle \bar\psi \psi \rangle$} \label{renormalization}

The classical stress-energy tensor in a FLRW spacetime has  two independent components. For a Dirac field, they are  (no sum on $i$),
\be T_0^0 = \frac{i}{2} \left( \bar{\psi} {\gamma}^0 \frac{\partial \psi}{\partial t} - \frac{\partial \bar{\psi}}{\partial t} {\gamma}^0 \psi \right) \ , \hspace{0.5cm}
T_i^i  = \frac{i }{2a} \left( \bar{\psi} {\gamma}^i \frac{\partial \psi}{\partial x^i} - \frac{\partial \bar{\psi}}{\partial x^i} {\gamma}^i \psi \right) \ . \label{ft00-ftii} \ee

We define the vacuum state $| 0 \rangle$ as $B_{\vec{k},\lambda} |0 \rangle \equiv D_{\vec{k},\lambda} |0 \rangle \equiv 0$, and denote any expectation value on this vacuum as e.g.~$\langle T_{\mu \nu} \rangle \equiv \langle 0 | T_{\mu \nu} | 0 \rangle$. In the quantum theory, the vacuum expectation values  of the stress-energy tensor take the form (see for example \cite{RNT})
\be
\left< T_{00}\right>=\frac{1}{2\pi^2 a^3}\int_{0}^{\infty} dk k^2 \rho_k (t)  \ , \hspace{0.5cm} \rho_k (t) \equiv 2 i \left( h_k^{I} \frac{\partial h_k^{I*}}{\partial t} + h_k^{II} \frac{\partial h_k^{II*}}{\partial t} \right) \ , \label{12}
\ee
and
\be \left< T_{ii}\right>=\frac{1}{2\pi^2 a}\int_{0}^{\infty} dk k^2 p_k (t)  \ , \hspace{0.5cm}  
p_k (t)\equiv-\frac{2k }{3a} ( h_k^{I} h_k^{II*}+h_k^{I*} h_k^{II} )  \ . \label{13}
\ee
The above formal expressions contain quartic, quadratic, and logarithmic UV divergences, which turn out to be independent of the particular quantum state. These divergences are similar to those described in \cite{Utiyama-DeWitt}. To characterize them, one plugs in \eqref{12}-\eqref{13} the adiabatic expansion of $h_k^{I}$ and $h_k^{II}$, given in Eq. (\ref{ferm-ansatz}). We shall see that, in the presence of a Yukawa interaction, all adiabatic orders up to the fourth one generate UV divergences. This is different to what happens in the case of a free field, where the divergences only appear at zeroth and second adiabatic orders \cite{RNT}. In general, adiabatic renormalization proceeds by subtracting those adiabatic terms from the integrand of the expectation values, producing a formal finite quantity. There are two important considerations regarding these subtractions. First, they must refer to all contributions of a given adiabatic term of fixed (adiabatic) order, othe
 rwise general covariance is not maintained. And second, one subtracts only the minimum number of terms required to get a finite result \cite{parker-toms}. 

We now proceed to calculate the renormalized expressions for the energy density and pressure. 

\subsection{Renormalized energy density}

We start by performing the adiabatic expansion of the energy density in momentum space (\ref{13})
\be \rho_k = \rho_k^{(0)} + \rho_k^{(1)} + \rho_k^{(2)} + \rho_k^{(3)} + \rho_k^{(4)} + \dots   \ , \ee
where $\rho_k^{(n)}$ is of $n$th adiabatic order. The adiabatic terms producing UV divergences (after integration in momenta) are
\bea \rho_k^{(0)} &=&-2 \omega \ , \label{rho0} \\
 \rho_k^{(1)} &=& -\frac{2 m s}{\omega } \ , \\
\rho_k^{(2)} &=&-\frac{\dot{a}^2 m^4}{4 a^2 \omega ^5}+\frac{\dot{a}^2 m^2}{4 a^2 \omega ^3}+\frac{m^2 s^2}{\omega ^3}-\frac{s^2}{\omega } \ ,  \\
\rho_k^{(3)}&=&\frac{5 \dot{a}^2 m^5 s}{4 a^2 \omega ^7}-\frac{7 \dot{a}^2 m^3 s}{4 a^2 \omega ^5}+\frac{\dot{a}^2 m s}{2 a^2 \omega ^3}-\frac{\dot{a} m^3
   \dot{s}}{2 a \omega ^5}+\frac{\dot{a} m \dot{s}}{2 a \omega ^3}-\frac{m^3 s^3}{\omega ^5}+\frac{m s^3}{\omega ^3} \ , \\
\rho_k^{(4)}  &=& \frac{105 \dot{a}^4 m^8}{64 a^4 \omega ^{11}}-\frac{91 \dot{a}^4 m^6}{32 a^4 \omega ^9}+\frac{81 \dot{a}^4 m^4}{64 a^4 \omega ^7}-\frac{\dot{a}^4
   m^2}{16 a^4 \omega ^5}-\frac{7 \dot{a}^2 m^6 \ddot{a}}{8 a^3 \omega ^9}+\frac{5 \dot{a}^2 m^4 \ddot{a}}{4 a^3 \omega ^7}-\frac{3 \dot{a}^2 m^2
   \ddot{a}}{8 a^3 \omega ^5}\\* && -\frac{35 \dot{a}^2 m^6 s^2}{8 a^2 \omega ^9}+\frac{15 \dot{a}^2 m^4 s^2}{2 a^2 \omega ^7}-\frac{m^4 \ddot{a}^2}{16 a^2
   \omega ^7}-\frac{27 \dot{a}^2 m^2 s^2}{8 a^2 \omega ^5}+\frac{m^2 \ddot{a}^2}{16 a^2 \omega ^5}+\frac{\dot{a}^2 s^2}{4 a^2 \omega
   ^3}+\frac{\dot{a} m^4 a^{(3)}}{8 a^2 \omega ^7}\nonumber \\* &&-\frac{\dot{a} m^2 a^{(3)}}{8 a^2 \omega ^5}+\frac{5 \dot{a} m^4 s \dot{s}}{2 a \omega
   ^7}-\frac{3 \dot{a} m^2 s \dot{s}}{a \omega ^5}+\frac{\dot{a} s \dot{s}}{2 a \omega ^3}+\frac{5 m^4 s^4}{4 \omega ^7}-\frac{3 m^2 s^4}{2 \omega
   ^5}-\frac{m^2 \dot{s}^2}{4 \omega ^5}+\frac{s^4}{4 \omega ^3}+\frac{\dot{s}^2}{4 \omega ^3} \nonumber \ , \eea
where we have used the notation $a^{(3)} \equiv d^3 a / dt^3  ,  a^{(4)} \equiv  d^4 a / dt^4  $, etc. 

We note that if we turn off the Yukawa coupling, we recover the results obtained in \cite{RNT}. The Yukawa interaction produces new contributions and, in particular, we have now non-zero terms at first and third adiabatic orders. The physical meaning of them will be given later on. Note here that in the UV limit, $\rho_k^{(0)} \sim k$, $(\rho_k^{(1)} + \rho_k^{(2)}) \sim k^{-1}$, and $(\rho_k^{(3)} + \rho_k^{(4)}) \sim k^{-3}$. This indicates that subtracting the zeroth-order term will cancel the natural quartic divergence of the stress-energy tensor, subtracting up to second order will cancel also the quadratic divergence, and subtracting up to fourth order will cancel the logarithmic divergence. Therefore, defining the adiabatic subtraction terms as
\be \label{Adsubtractions}\langle T_{00} \rangle_{Ad} \equiv \frac{1}{2 \pi^2 a^3} \int_0^{\infty} dk k^2 (\rho_k^{(0)}+\rho_k^{(1)}+ \rho_k^{(2)} + \rho_k^{(3)}+\rho_k^{(4)} )\equiv \frac{1}{2 \pi^2 a^3} \int_0^{\infty} dk k^2 \rho_k^{(0-4)} \ , \ee
the renormalized 00 component of the stress-energy tensor is 
\be \langle T_{00} \rangle_{ren} \equiv   \langle T_{00} \rangle - \langle T_{00} \rangle_{Ad}=      \frac{1}{2 \pi^2 a^3} \int_0^{\infty} dk k^2 ( \rho_k - \rho_k^{(0-4)}) \ . \label{ren-ten} \ee
This integral is, by construction, finite.

\subsection{Renormalized pressure}

The method proceeds in the same way for the pressure. The renormalized ii component of the stress-energy tensor is given by
\bea
\left< T_{ii}\right>_{ren}\equiv \left< T_{ii}\right> - \left< T_{ii}\right>_{Ad}= \frac{1}{2\pi^2 a}\int_{0}^{\infty} dk k^2 (p_k-p_k^{(0-4)}) \ ,  \label{17b}
\eea
where $p_k^{(0-4)}\equiv  p_k^{(0)}+p_k^{(1)}+ p_k^{(2)}+p_k^{(3)}+p_k^{(4)}$, and
\be \left< T_{ii}\right>_{Ad}\equiv  \frac{1}{2\pi^2 a}\int_{0}^{\infty} dk k^2 p_k^{(0-4)}    \label{iiad} \ . \ee
The corresponding adiabatic terms for the pressure are
\bea
p_k^{(0)} & = &-\frac{2 \omega }{3} + \frac{2 m^2}{3 \omega},  \label{17c}\\
p_k^{(1)} & = & \frac{2 m s}{3 \omega }-\frac{2 m^3 s}{3 \omega ^3}, \\
p_k^{(2)} & = &-\frac{5 \dot{a}^2 m^6}{12 a^2 \omega ^7}+\frac{\dot{a}^2 m^4}{2 a^2 \omega ^5}-\frac{\dot{a}^2 m^2}{12 a^2 \omega ^3}+\frac{m^4 \ddot{a}}{6 a
   \omega ^5}-\frac{m^2 \ddot{a}}{6 a \omega ^3}+\frac{m^4 s^2}{\omega ^5}-\frac{4 m^2 s^2}{3 \omega ^3}+\frac{s^2}{3 \omega },   \label{17d}\\
p_k^{(3)}& = &-\frac{35 \dot{a}^2 m^7 s}{12 a^2 \omega ^9}-\frac{5 \dot{a}^2 m^5 s}{a^2 \omega ^7}+\frac{9 \dot{a}^2 m^3 s}{4 a^2 \omega ^5}-\frac{\dot{a}^2 m s}{6
   a^2 \omega ^3}-\frac{5 m^5 s \ddot{a}}{6 a \omega ^7}-\frac{5 \dot{a} m^5 \dot{s}}{6 a \omega ^7}+\frac{7 m^3 s \ddot{a}}{6 a \omega ^5}+\frac{7
   \dot{a} m^3 \dot{s}}{6 a \omega ^5} \nonumber \\* && -\frac{m s \ddot{a}}{3 a \omega ^3}-\frac{\dot{a} m \dot{s}}{3 a \omega ^3}-\frac{5 m^5 s^3}{3 \omega
   ^7}+\frac{8 m^3 s^3}{3 \omega ^5}+\frac{m^3 \ddot{s}}{6 \omega ^5}-\frac{m s^3}{\omega ^3}-\frac{m \ddot{s}}{6 \omega ^3} \ ,  \\
p_k^{(4)} & = &\frac{385 \dot{a}^4 m^{10}}{64 a^4 \omega ^{13}}-\frac{791 \dot{a}^4 m^8}{64 a^4 \omega ^{11}}+\frac{1477 \dot{a}^4 m^6}{192 a^4 \omega
   ^9}-\frac{m^4 a^{(4)}}{24 a \omega ^7}-\frac{263 \dot{a}^4 m^4}{192 a^4 \omega ^7}+\frac{m^2 a^{(4)}}{24 a \omega ^5}+\frac{\dot{a}^4
   m^2}{48 a^4 \omega ^5}\nonumber \\* &&-\frac{77 \dot{a}^2 m^8 \ddot{a}}{16 a^3 \omega ^{11}}+\frac{203 \dot{a}^2 m^6 \ddot{a}}{24 a^3 \omega ^9}-\frac{191
   \dot{a}^2 m^4 \ddot{a}}{48 a^3 \omega ^7}+\frac{\dot{a}^2 m^2 \ddot{a}}{3 a^3 \omega ^5}-\frac{105 \dot{a}^2 m^8 s^2}{8 a^2 \omega
   ^{11}}+\frac{665 \dot{a}^2 m^6 s^2}{24 a^2 \omega ^9}+\frac{7 m^6 \ddot{a}^2}{16 a^2 \omega ^9}\nonumber \\ &&-\frac{145 \dot{a}^2 m^4 s^2}{8 a^2 \omega
   ^7}-\frac{5 m^4 \ddot{a}^2}{8 a^2 \omega ^7}+\frac{29 \dot{a}^2 m^2 s^2}{8 a^2 \omega ^5}+\frac{3 m^2 \ddot{a}^2}{16 a^2 \omega
   ^5}-\frac{\dot{a}^2 s^2}{12 a^2 \omega ^3}+\frac{7 \dot{a} m^6 a^{(3)}}{12 a^2 \omega ^9}-\frac{5 \dot{a} m^4 a^{(3)}}{6 a^2 \omega
   ^7}\nonumber \\ &&+\frac{\dot{a} m^2 a^{(3)}}{4 a^2 \omega ^5}+\frac{35 m^6 s^2 \ddot{a}}{12 a \omega ^9}+\frac{35 \dot{a} m^6 s \dot{s}}{6 a \omega
   ^9}-\frac{5 m^4 s^2 \ddot{a}}{a \omega ^7}-\frac{10 \dot{a} m^4 s \dot{s}}{a \omega ^7}+\frac{9 m^2 s^2 \ddot{a}}{4 a \omega ^5}+\frac{9 \dot{a}
   m^2 s \dot{s}}{2 a \omega ^5}\nonumber \\ &&-\frac{s^2 \ddot{a}}{6 a \omega ^3}-\frac{\dot{a} s \dot{s}}{3 a \omega ^3}+\frac{35 m^6 s^4}{12 \omega ^9}-\frac{65
   m^4 s^4}{12 \omega ^7}-\frac{5 m^4 s \ddot{s}}{6 \omega ^7}-\frac{5 m^4 \dot{s}^2}{12 \omega ^7}+\frac{11 m^2 s^4}{4 \omega ^5}+\frac{m^2 s
   \ddot{s}}{\omega ^5}+\frac{m^2 \dot{s}^2}{3 \omega ^5}\nonumber \\ &&-\frac{s^4}{4 \omega ^3}-\frac{s \ddot{s}}{6 \omega ^3}+\frac{\dot{s}^2}{12 \omega ^3}\ .
\eea
As before, we see that in the UV limit, $p_k^{(0)} \sim k$, $(p_k^{(1)} + p_k^{(2)} ) \sim k^{-1}$, and $(p_k^{(3)} + p_k^{(4)} ) \sim k^{-3}$. Subtracting the zeroth-order term eliminates the quartic divergence, subtracting up to second order removes the quadratic divergence, and subtracting up to fourth order removes the logarithmic divergence. If the Yukawa interaction is removed, we recover again the results in \cite{RNT}. The interaction produces also nonzero contributions to the first and third adiabatic orders.

\subsection{Renormalization of $\langle \bar{\psi} \psi\rangle$}
 
  We are also interested in computing the renormalized expectation value $\langle \bar{\psi} \psi\rangle_{ren}$. The formal (unrenormalized) expression for this quantity is
 \be \langle \bar{\psi} \psi\rangle = \frac{-1}{\pi^2a^3}\int_0^{\infty} dk k^2 \langle \bar{\psi} \psi\rangle_k  \ , \hspace{0.5cm} \langle \bar{\psi} \psi\rangle_k \equiv |h_k^{I}|^2 - |h_k^{II}|^2 \ . \label{eqbilineal1} \ee  
  We define the corresponding terms in the adiabatic expansion as $ \langle \bar{\psi} \psi\rangle_k = \langle \bar{\psi} \psi\rangle_k^{(0)} + \langle \bar{\psi} \psi\rangle_k^{(1)}+\langle \bar{\psi} \psi\rangle_k^{(2)}+\langle \bar{\psi} \psi\rangle_k^{(3)}+ ...$\ . Due to the Yukawa interaction, ultraviolet divergences arrive till the third adiabatic order. In general, we have
\bea
 \langle \bar{\psi} \psi \rangle_k^{(n)} =\frac{\omega+m}{2 \omega}\left(|F|^2\right)^{(n)}-\frac{\omega-m}{2\omega}\left(|G|^2\right)^{(n)}  \label{eqbilineal2}\ .
\eea
From here, we obtain 
\bea \langle \bar{\psi} \psi\rangle_k^{(0)} &=& \frac{m}{\omega } \ , \\
\langle \bar{\psi} \psi\rangle_k^{(1)} &=& \frac{s}{\omega }-\frac{m^2 s}{\omega ^3} \ , \\
\langle \bar{\psi} \psi\rangle_k^{(2)} &=& -\frac{5 \dot{a}^2 m^5}{8 a^2 \omega ^7}+\frac{7 \dot{a}^2 m^3}{8 a^2 \omega
   ^5}-\frac{\dot{a}^2 m}{4 a^2 \omega ^3}+\frac{m^3 \ddot{a}}{4 a \omega ^5}-\frac{m
   \ddot{a}}{4 a \omega ^3}+\frac{3 m^3 s^2}{2 \omega ^5}-\frac{3 m s^2}{2 \omega ^3} \ , \\
\langle \bar{\psi} \psi\rangle_k^{(3)} &=&\frac{35 \dot{a}^2 m^6 s}{8 a^2 \omega ^9}-\frac{15 \dot{a}^2 m^4 s}{2 a^2 \omega
   ^7}+\frac{27 \dot{a}^2 m^2 s}{8 a^2 \omega ^5}-\frac{\dot{a}^2 s}{4 a^2 \omega
   ^3}-\frac{5 m^4 s \ddot{a}}{4 a \omega ^7}-\frac{5 \dot{a} m^4 \dot{s}}{4 a \omega
   ^7}+\frac{3 m^2 s \ddot{a}}{2 a \omega ^5}+\frac{2 \dot{a} m^2 \dot{s}}{a \omega
   ^5}\nonumber \\ &&-\frac{s \ddot{a}}{4 a \omega ^3}-\frac{3 \dot{a} \dot{s}}{4 a \omega ^3}-\frac{5
   m^4 s^3}{2 \omega ^7}+\frac{3 m^2 s^3}{\omega ^5}+\frac{m^2 \ddot{s}}{4 \omega
   ^5}-\frac{s^3}{2 \omega ^3}-\frac{\ddot{s}}{4 \omega ^3} \ . \eea
The adiabatic prescription leads then to 
 \be \langle \bar{\psi} \psi\rangle_{ren}=\langle \bar{\psi} \psi\rangle - \langle \bar{\psi} \psi\rangle_{Ad}=  \frac{-1}{\pi^2a^3}\int_0^{\infty} dk k^2 ( \langle \bar{\psi} \psi\rangle_k-\langle \bar{\psi} \psi\rangle_k^{(0-3)}) \ . \label{eq:ren-variance} \ee
 
In this case, we observe that in the UV limit, $(\langle \bar{\psi} \psi\rangle_k^{(0)} + \langle \bar{\psi} \psi\rangle_k^{(1)} )\sim k^{-1}$, and  $(\langle \bar{\psi} \psi\rangle_k^{(2)} + \langle \bar{\psi} \psi\rangle_k^{(3)} )\sim k^{-3} )$. Subtracting up to first order eliminates the quadratic divergence, and up to third order removes the logarithmic one.

 Our results can be generically  implemented together with numerical methods to compute the renormalized expectation values $\langle T_{\mu\nu} \rangle_{ren}$ and $\langle \bar\psi\psi\rangle_{ren}$. On the other hand, we would like to briefly  comment that a higher-order adiabatic expansion  also serves to generate asymptotic analytical expressions for the renormalized stress-energy tensor in some special situations. This happens in spacetime regions where the relevant  modes always evolve adiabatically. For instance, if we  approximate the form of the exact modes $\{ h_k^{I}, h_k^{II} \}$ by their higher-order adiabatic expansion, we can find in a very straightforward way an analytic approximation for the renormalized quantities in the adiabatic regime,  as in the example given in Appendix B.  Outside the adiabatic regime one should use numerical methods to find the exact modes and plug them in the generic renormalized expressions obtained above.

%{\color {red} Our results can be generically used to implement numerical methods to compute the renormalized expectation values $\langle T_{\mu\nu} \rangle_{ren}$ and $\langle \bar\psi\psi\rangle_{ren}$. However, we would like to briefly  comment that a higher-order adiabatic expansion  also serves to generate asymptotic analytical expressions for the renormalized stress-energy tensor in some special situations. This happens in spacetime regions where the relevant  modes always evolve adiabatically. For instance, if we  approximate the form of the exact modes $\{ h_k^{I}, h_k^{II} \}$ by their higher-order adiabatic expansion, we can find in a very straightforward way an analytic approximation for the renormalized quantities in the adiabatic regime,  as in the example given in Appendix B.  Outside the adiabatic regime one should use numerical methods to find the exact modes and plug them in the generic renormalized expressions obtained above.}

%In a spacetime region where the modes evolve adiabatically, the adiabatic expansion  can approximate well the   relevant  modes contributing to the bulk of the renormalized stress-energy tensor or the two-point function.}

%In a more general scenario, our results can be used to implement numerical methods to compute the renormalized expectation values $\langle T_{\mu\nu} \rangle_{ren}$ and $\langle \bar\psi\psi\rangle_{ren}$. We shall see an example of this in Appendix \ref{appendixB}.

\section{Ultraviolet divergences and renormalization counterterms}\label{sectioncounterterms}

The ultraviolet divergent terms of the adiabatic subtractions can be univocally related to particular counterterms in a Lagrangian density including the background gravity-scalar sector.
By writing
\bea \mathcal L &=& \mathcal L_m + {\sqrt{-g}}\left[ \frac{1}{2}g^{\mu\nu}\nabla_\mu \Phi \nabla_\nu \Phi -\sum_{i=1}^4  \frac{\lambda_i }{i!}\Phi^i -\xi_1 R\Phi-\frac{1}{2}\xi_2R \Phi^2 - \frac{1}{8\pi G}\Lambda + \frac{1}{16\pi G}R\right] \nonumber \\
&+& {\sqrt{-g}}\left[ \frac{1}{2}\delta Zg^{\mu\nu}\nabla_\mu \Phi \nabla_\nu \Phi -\sum_{i=1}^4  \frac{\delta {\lambda_i} }{i!}\Phi^i -\delta {\xi_1}R\Phi-\frac{1}{2}\delta {\xi_2}R \Phi^2 - \frac{1}{8\pi }\delta {\Lambda} + \frac{1}{16\pi }\delta {G^{-1}}R\right]\ , \eea 
the equations of motion for the scalar field are
\bea \label{eqPhi3}&&(1+\delta Z) \Box \Phi +(\lambda_1+\delta {\lambda_1})+(\lambda_2+\delta {\lambda_2}) \Phi + (\lambda_3+\delta {\lambda_3})\frac{1}{2}\Phi^2 \nonumber \\ &+& \frac{1}{3!}(\lambda_4+ \delta {\lambda_4})\Phi^3+ (\xi_1+\delta {\xi_1})R+(\xi_2+\delta {\xi_2})R\Phi  = - g_Y \langle \bar \psi \psi \rangle\ . \eea
From (\ref{eq:ren-variance}), we can write the identity
\be \label{eqmc}\langle \bar \psi \psi \rangle=\langle \bar \psi \psi \rangle_{ren}  + \frac{1}{\pi^2a^3}\int_0^{\infty} dk k^2 (\langle \bar{\psi} \psi\rangle_k^{(0)}+ \langle \bar{\psi} \psi\rangle_k^{(1)}+ \langle \bar{\psi} \psi\rangle_k^{(2)}+\langle \bar{\psi} \psi\rangle_k^{(3)}) \ ,\ee
where $\langle \bar \psi \psi \rangle_{ren}$ is finite and the remaining integrals at the right-hand side  of (\ref{eqmc}) are the  adiabatic subtraction terms.
As we shall see, the ultraviolet divergences of the adiabatic subtraction terms can be removed by counterterms of the form: $\delta {Z}\Box\Phi$, $\delta {\lambda_1}$, 
$\delta {\lambda_2} \Phi$,  $\delta {\lambda_3}\Phi^2$,  $\delta {\lambda_4}\Phi^3$, $\delta {\xi_1}R$, and $\delta {\xi_2}R\Phi$. To deal with the UV-divergent subtraction terms  we use dimensional regularization \cite{Bunch80}. We can check that the (covariantly) regulated divergences take the same form as the above covariant counterterms. For $\langle \bar{\psi}\psi \rangle^{(0)}$ we have ($n$ denotes the spacetime dimension)
\be
 \langle \bar{\psi}\psi \rangle^{(0)}=-\frac{1}{\pi^2a^3}\int^{\infty}_{0} dk k^{2}\left(-\frac{m}{\omega(t)}\right)\to -\frac{1}{\pi^2a^3}\int^{\infty}_{0} dk k^{n-2}\left(-\frac{m}{\omega(t)}\right)=\frac{m^3}{2\pi^2(n-4)} + ... 
 \ee
 where we will retain only the poles at $n=4$.
This divergence can be absorbed by $\delta \lambda_1$. Additionally, we also have
\be
 \langle \bar{\psi}\psi \rangle^{(1)}=-\frac{1}{\pi^2a^3}\int^{\infty}_{0} dk k^{n-2}\left(- \frac{s(t) k^2}{\omega^3(t) a(t)^2}\right)=\frac{3 g_Y m^2}{2\pi^2(n-4)}\Phi(t)+\cdots \ . 
 \ee
This divergence of adiabatic order one can be absorbed by $\delta \lambda_2$.  The divergences of adiabatic order two
\be \langle \bar{\psi}\psi \rangle^{(2)}=-\frac{m}{24\pi^2(n-4)}R+\frac{3mg_Y^2}{2\pi^2(n-4)}\Phi^2(t)+\cdots \ee
can also be eliminated by $\delta \xi_1$ and $\delta \lambda_3$. Finally, the three divergences of adiabatic order three
\be \langle \bar{\psi}\psi \rangle^{(3)}=\frac{g_Y}{4\pi^2(n-4)}\Box\Phi(t) +\frac{g_Y}{24 \pi^2 (n-4)}R\Phi(t)+\frac{g_Y^3}{2\pi^2 (n-4)}\Phi^3(t)+\cdots
\ee
are absorbed by $\delta Z$, $\delta \xi_2$ and $\delta \lambda_4$. 

On the other hand, the tensorial equations are
\bea \label{eqg2} && \frac{1}{8\pi} \left( \frac{1}{G}+\delta {G^{-1}} \right) G^{\mu\nu} + \frac{1}{8\pi} \left( \frac{\Lambda}{G}+\delta {\Lambda} \right) g^{\mu\nu}+ (1+\delta Z) ( \nabla^\mu\Phi \nabla^\nu\Phi - \frac{1}{2}g^{\mu\nu} \nabla^\rho\Phi \nabla_\rho\Phi )  \nonumber \\
 &+& g^{\mu\nu}\sum_{i=1}^4  \frac{(\lambda_i+\delta {\lambda_i}) }{i!}\Phi^i -2\sum_{i=1}^2 \frac{\xi_i + \delta \xi_i}{i!}  (G^{\mu\nu}\Phi^i -g^{\mu\nu}\Box\Phi^i+\nabla^{\mu}
  \nabla^{\nu}\Phi^i) = - \langle T_m^{\mu\nu}\rangle \ ,  \eea
and we find similar cancellations.  However, two extra divergences appear. Focusing, for simplicity, at zeroth adiabatic order, we have
\bea -\frac{1}{2 \pi^2 a^3} \int_0^{\infty} dk k^{n-2}  \rho_k^{(0)} &\approx&\frac{m^4}{8 \pi^2} \frac{1}{n-4} \ , \nonumber \\
-\frac{1}{2 \pi^2 a^3} \int_0^{\infty} dk k^{n-2}  a^2 p_k^{(0)}&\approx&-\frac{m^4a^2}{8 \pi^2} \frac{1}{n-4} \ . \eea
At first adiabatic order we encounter the following divergences  
\bea 
 -\frac{1}{2 \pi^2 a^3} \int_0^{\infty} dk k^{n-2}  \rho_k^{(1)}&\approx&\frac{m^3 s}{2 \pi^2} \frac{1}{n-4} \ , \nonumber \\
-\frac{1}{2 \pi^2 a^3} \int_0^{\infty} dk k^{n-2}  a^2 p_k^{(1)} &\approx& -\frac{m^3 a^2 s}{2 \pi^2} \frac{1}{n-4} 
  \ .  \label{ii1}
   \eea
At second adiabatic order we find these divergences
\bea  -\frac{1}{2 \pi^2 a^3} \int_0^{\infty} dk k^{n-2}  \rho_k^{(2)}&\approx&\frac{m^2}{8 \pi^2} \frac{1}{n-4}\frac{\dot a^2}{a^2} + \frac{3m^2}{ 4\pi^2} \frac{1}{n-4} s^2  \ , \nonumber \\
-\frac{1}{2 \pi^2 a^3} \int_0^{\infty} dk k^{n-2}  a^2 p_k^{(2)} &\approx& -\frac{m^2 a^2}{24 \pi^2} \frac{1}{n-4} \left(2\frac{\ddot a}{a}+\frac{\dot a^2}{a^2}\right)  -\frac{3a^2 m^2 s^2 }{4 \pi^2} \frac{1}{n-4}
  \ .  \label{ii2} \eea
At third adiabatic order we get the following divergences ($H\equiv \dot a/a$)
\bea 
 -\frac{1}{2 \pi^2 a^3} \int_0^{\infty} dk k^{n-2}  \rho_k^{(3)}&\approx&\frac{m}{12 \pi^2} \frac{1}{n-4}\left[3H^2s+3H\dot s+6s^3 \right] \ ,  \label{ii3} \\
-\frac{1}{2 \pi^2 a^3} \int_0^{\infty} dk k^{n-2}  a^2 p_k^{(3)} &\approx& -\frac{m a^2}{12 \pi^2} \frac{1}{n-4} \left[\ddot s+2H\dot s+\left(H^2+2\frac{\ddot a}{a} \right)s+6s^3 \right]   \nonumber
  \ . 
   \eea
Finally, at fourth adiabatic order the divergences are
\bea 
 -\frac{1}{2 \pi^2 a^3} \int_0^{\infty} dk k^{n-2}  \rho_k^{(4)}&\approx&\frac{1}{8 \pi^2} \frac{1}{n-4}\left[H^2s^2+s^4+2H\dot s s +\dot s^2 \right]   \ , \label{ii4}\\
-\frac{1}{2 \pi^2 a^3} \int_0^{\infty} dk k^{n-2}  a^2 p_k^{(4)} &\approx& -\frac{ a^2}{8 \pi^2} \frac{1}{n-4} \left[s^4+\left(H^2+2\frac{\ddot a}{a} \right)\frac{s^2}{3}-\frac{\dot s^2}{3}+\frac{4}{3}H s \dot s+\frac{2}{3}s \ddot s \right]  
  \ .  \nonumber
   \eea
All the above divergent expressions arising from the Yukawa interaction can be written covariantly as
\bea
\langle T_{\mu\nu}\rangle_{Ad}^{(0)} & \approx & \frac{m^4}{8\pi^2(n-4)}g_{\mu\nu} \ , \\
\langle T_{\mu\nu} \rangle_{Ad}^{(1)} & \approx & \frac{g_Y \Phi m^3}{2\pi^2(n-4)}g_{\mu\nu} \ ,  \\
\langle T_{\mu\nu} \rangle_{Ad}^{(2)} & \approx & \frac{3g_Y^2\Phi^2 m^2}{4\pi^2(n-4)}g_{\mu\nu} - \frac{m^2}{24\pi^2(n-4)} G_{\mu\nu}  \ , \\
\langle T_{\mu\nu}\rangle_{Ad}^{(3)} & \approx & -\frac {m g_Y}{12\pi^2(n-4)}\left[ G_{\mu\nu}\Phi-\Box \Phi g_{\mu\nu}+\nabla_{\mu}\nabla_{\nu}\Phi-6g_Y^2 \Phi^3 g_{\mu\nu} \right]  \ ,\\
\langle T_{\mu\nu}\rangle_{Ad}^{(4)} & \approx & \frac {-g_Y^2}{24\pi^2(n-4)}\left[G_{\mu\nu}\Phi^2- g_{\mu\nu}\Box \Phi^2+\nabla_{\mu}\nabla_{\nu}\Phi^2-6(\nabla_\mu\Phi \nabla_\nu\Phi - \frac{1}{2}g_{\mu\nu} \nabla_\rho\Phi \nabla^\rho\Phi)-3 g_Y^2 \Phi^4 g_{\mu\nu}  \right]  \ , \ \ \ \ \ \ 
\eea
and can be consistently removed (including also the divergences for  $\langle \bar \psi\psi\rangle$)  by the renormalization parameters
  \be \delta \Lambda =  -\frac{m^4}{\pi(n-4)} \ , \hspace{0.4cm}  \delta G^{-1} =\frac{m^2}{3\pi(n-4)} \ , \hspace{0.4cm}   \delta Z  =  -\frac{g_Y^2}{4\pi^2(n-4)} \ , \ee
 \be \delta \lambda_1 =  -\frac{m^3g_Y}{2\pi^2(n-4)} \ , \hspace{0.4cm}  \delta \lambda_2 = -\frac{3m^2 g_Y^2}{2\pi^2(n-4)} \ , \hspace{0.4cm}  \delta \lambda_3 = -\frac{3 m g_Y^3}{\pi^2(n-4)} \ , \hspace{0.4cm}  \delta \lambda_4  =  -\frac{3 g_Y^4}{\pi^2(n-4)} \ , \ee
\be \delta \xi_1  =  -\frac{m g_Y}{24\pi^2(n-4)} \ , \hspace{0.4cm} \delta \xi_2  =  -\frac{g_Y^2}{24\pi^2(n-4)} \ . \ee 

 We remark that the set of needed counterterms is all possible counterterms having  couplings with non-negative mass dimension, up to Newton's coupling constant. This is also in agreement with the results in perturbative Quantum Field Theory in flat spacetime. The renormalizability of the Yukawa interaction $g_Y\varphi\bar\psi \psi$ of a quantized massive scalar field $\varphi$ with a massive quantized Dirac field $\psi$ requires us to add terms of the form $\frac{\lambda Z_{\lambda}}{4!}\varphi^4$, $\frac{\kappa Z_{\kappa}}{3!}\varphi^3$, and also a term linear in $\varphi$ \cite{Srednicki}.  
 The presence of a curved background would require us to add the terms $\xi_1R \varphi$ and $\xi_2R \varphi^2$.  We note that a term of the form $\xi_2R \varphi^2$
 is required by renormalization for a purely quantized scalar field $\varphi$ if  a self-interaction term of the form $\frac{\lambda}{4!} \varphi^4$ appears in the bare Lagrangian density \cite{bunch-parker, bunch81}.  Here we have found that the Yukawa interaction demands the presence of the renormalized terms  $\xi_1R \varphi$ and $\xi_2R \varphi^2$ (as well as the terms $\lambda_i \varphi^i$), even if they are not present in the bare Lagrangian density. Similar counterterms have been identified in the approach  in Ref. \cite{Baacke-Patzold}. 
  
 Therefore, the  tentative semiclassical equations presented in Sec. \ref{semiclassicalequations} should  be reconsidered to  include the above-required counterterms. In terms of the renormalized parameters we have
 \bea \label{eqg3} && \frac{1}{8\pi G}(G^{\mu\nu} + \Lambda g^{\mu\nu})+ (\nabla^\mu\Phi \nabla^\nu\Phi - \frac{1}{2}g^{\mu\nu} \nabla^\rho\Phi \nabla_\rho\Phi + V(\Phi) g^{\mu\nu})  \nonumber \\
 &-&  2\sum_{i=1}^2 \frac{\xi_i}{i!} (G^{\mu\nu}\Phi^i -g^{\mu\nu}\Box\Phi^i+\nabla^{\mu}
  \nabla^{\nu}\Phi^i) = - \langle T_m^{\mu\nu}\rangle_{ren} \ ,  \eea
and 
  \be \label{eqPhi4}\Box \Phi +\frac{\partial V}{\partial \Phi} + \xi_1R+\xi_2R\Phi= - g_Y \langle \bar \psi \psi \rangle_{ren}\ , \ee
where the potential $V(\Phi)$ should contain the terms
\be V(\Phi) = \lambda_1\Phi + \frac{\lambda_2}{2}\Phi^2 + \frac{\lambda_3}{3!}\Phi^3 + \frac{\lambda_4}{4!}\Phi^4  \ . \ee
Obviously, additional terms, not required by renormalization, can be added to the potential if one adopts an effective field theory viewpoint. Some of the renormalized parameters ($\Lambda$, $\xi_1$, $\lambda_1$, $\cdots$) could take, by fine-tuning, zero values. We do not consider these issues in this work.
  
  \section{Conformal anomaly}\label{sec:conformalanomaly}
  
  In this section we will analyze the massless limit of the theory and work out the conformal anomaly. In the massless limit the classical action of the theory enjoys invariance under the conformal transformations
\be  g_{\mu\nu}(x) \to \Omega^2(x)g_{\mu\nu}(x) \ , \  \ \ \  \Phi(x) \to \Omega^{-1}(x) \Phi(x) \ , \ee
with 
\be \psi (x) \to \Omega^{-3/2}(x) \psi(x) \ , \ \ \ \ \  \bar \psi (x) \to \Omega^{-3/2}(x) \bar \psi(x). \ee
Variation of the action yields the identity
\be g^{\mu\nu}T_{\mu\nu}^m + \Phi \frac{1}{\sqrt{-g}}\frac{\delta S_m}{\delta \Phi} =0 \ , \ee
which, in our case, turns out to be $ g^{\mu\nu} T_{\mu\nu} -g_Y \Phi \bar\psi\psi = 0$. At the quantum level the theory will lose its conformal invariance  as a consequence of  renormalization [which respects general covariance and hence (\ref{covariance})] and generates an anomaly
\be g^{\mu\nu} \langle T_{\mu\nu}^m\rangle_{ren}  -g_Y \Phi \langle \bar\psi\psi\rangle_{ren} = C_f \neq 0 \ . \ee
$C_f$ is independent of the quantum state and depends only on local  quantities of the external fields. 

To  calculate the conformal anomaly in the adiabatic regularization method, we have to start with a massive field and take the massless limit at the end of the calculation. Therefore, 
\be C_f= g^{\mu\nu} \langle T_{\mu\nu}^m\rangle_{ren}  -g_Y\Phi\langle \bar \psi \psi  \rangle_{ren} = \lim_{m\to 0} m ( \langle \bar \psi \psi\rangle_{ren}-\langle \bar \psi \psi\rangle^{(4)}) \ . \ee
Since the divergences of the stress-energy tensor have terms of fourth adiabatic order, the adiabatic subtractions for  $\langle \bar \psi \psi\rangle$ should also include them. The fourth-order subtraction term, which produces a nonzero finite contribution when $m\to 0$, is codified in $\langle \bar \psi \psi\rangle^{(4)}$. The term $m \langle \bar \psi \psi\rangle_{ren}$ vanishes when $m \to 0$. The remaining piece produces the anomaly [recall (\ref{eqbilineal1})-(\ref{eqbilineal2})]
 \bea C_f
&=&-\lim_{m \to 0} \frac{m}{\pi^2a^3} 
 \int^{\infty}_{0}dk k^2 \left( -\frac{(\omega +m)}{2 \omega}[ F^{(4)} + F^{(4)*} + F^{(1)}F^{(3)*} + F^{(1)*}F^{(3)} + |F^{(2)}|^2] \right. \nonumber \\
&&+ \left. \frac{(\omega -m)}{2 \omega}[ G^{(4)} + G^{(4)*} + G^{(1)}G^{(3)*} + G^{(1)*}G^{(3)} + |G^{(2)}|^2] \right) \ . \eea 
Applying the adiabatic expansion computed in Sec. \ref{adiabaticexpansion} and doing the integrals we obtain 
\bea \label{Cf}C_f=\frac{a^{(4)}}{80\pi^2 
a}+\frac{s^2\ddot{a}}{8\pi^2a}+\frac{\ddot{a}^2}{80\pi^2a}+\frac{3s\dot{s}\dot{a}}{4\pi^2a}+\frac{s^2\dot{a}^2}{8\pi^2a^2}+\frac{3\dot{a}a^{(3)}}{80\pi^2a^2}-\frac{\dot{a}^2\ddot{a}}{60\pi^2a^3}+\frac{s\ddot{s}}{4\pi^2}+\frac{\dot{s}^2}{8\pi^2}+\frac{s^4}{8\pi^2}
\ . \ \ \ \ \ \eea
Since $C_f$ is a scalar, we must be able to rewrite the above result as a linear combination of covariant scalar terms made out of the metric, the Riemann tensor, covariant derivatives, and the external scalar field $\Phi$. Our result is
\bea \label{Cf2}C_f=\frac{1}{2880\pi^2}\left[-11\left(R_{\alpha \beta}R^{\alpha \beta}-\frac13 R^2\right)+6\Box 
R\right]+\frac{g_Y^2}{8\pi^2}\left[\nabla^{\mu}\Phi\nabla_{\mu}\Phi+2\Phi\Box \Phi +\frac16 \Phi^2 R + g_Y^2\Phi^4
\right]. \label{traceanomaly} \ \ \ \ \ \ \ \ 
\eea
The same result is obtained by using the results of Sec. \ref{renormalization}.  $C_f$ can be reexpressed as [recall (\ref{ren-ten})-(\ref{17b})]
\bea
C_f= \lim_{m\to 0} \frac{-1}{2\pi^2 a^3}\int^{\infty}_{0}dk k^2 \left(\rho_k^{(0-4)}-3p_k^{(0-4)}-2(s(t)+m) \langle \bar{\psi} \psi \rangle_k^{(0-3)}\right)
\ . \eea
Performing the integrals we get exactly (\ref{Cf}) and hence (\ref{Cf2}). 

In Appendix \ref{appendixA} we have computed the conformal anomaly for a massless scalar field $\phi$ with conformal coupling to the scalar curvature $\xi=1/6$, and with a Yukawa-type interaction of the form $g_Y^2\Phi^2 \phi^2$. Adiabatic regularization predicts the following conformal anomaly
\be \label{ACs}
  C_s=\frac{1}{2880\pi^2}\left[ \Box R-\left
  (R^{\mu\nu}R_{\mu\nu}-\frac13R^2\right)\right] -\frac{h^2}{48\pi^2}(\Phi\Box \Phi +
  \nabla^{\mu}\Phi\nabla_{\mu}\Phi
  +\frac{3h^2}{2}\Phi^4) \ .
\ee

In the absence of Yukawa interaction ($h=0$, $g_Y=0$) we reproduce the well-known trace anomaly for both scalar and spin-$1/2$ fields  (restricted to our FLRW spacetime) \cite{birrell-davies}. We recall that the trace anomaly is generically given for a conformal free field of  spin $0, 1/2$, or $1$ in terms of three coefficients
\bea
&& g^{\mu\nu}\left\langle T_{\mu\nu}\right\rangle _{ren}=aC_{\mu\nu\rho\sigma}C^{\mu\nu\rho\sigma}+bG +c\square R \ , \,\,\, \label{eq:anomtrace}
\eea
where $C_{\mu\nu\rho\sigma}$ is the Weyl tensor and $G= R_{\mu\nu\rho\sigma}R^{\mu\nu\rho\sigma} - 4R_{\mu\nu}R^{\mu\nu} + R^2$ is proportional to the  Euler density. The coefficients $a$ and $b$ are independent of the renormalization scheme and are given by \cite{Duff, HHR}
\bea \label{Tranomaly}a&=& \frac{1}{120(4\pi)^2} (N_s+6N_f +12N_v) \ , \nonumber \\
b&=& \frac{-1}{360(4\pi)^2} (N_s+11N_f+62N_v) \ , \eea
where $N_s$ is the number of real scalar fields, $N_f$ is the number of Dirac fields, and $N_v$ is the number of vector fields.  Our results with $g_Y=0$ fit the values in (\ref{Tranomaly}). [We note that in the FLRW spacetime of adiabatic regularization the Weyl tensor  vanishes identically]. In contrast, the coefficient $c$ depends in general on the particular renormalization scheme \cite{Wald78}. A local counterterm proportional to $R^2$ in the action can modify the coefficient $c$. For instance, for vector fields the point splitting and the dimensional regularization method predict different values for $c$.

When the Yukawa interaction is added, the general form of the conformal anomaly is 
\bea
 g^{\mu\nu}\left\langle T_{\mu\nu}^m\right\rangle _{ren} + \Phi \frac{1}{\sqrt{-g}}\langle \frac{\delta S_m}{\delta \Phi}\rangle_{ren} &=&aC_{\mu\nu\rho\sigma}C^{\mu\nu\rho\sigma}+bG +c\square R  \nonumber \\ 
&+& d \,g_Y^2 \nabla^{\mu}\Phi\nabla_{\mu}\Phi+ e \, g_Y^2 \Phi\Box \Phi + f g_Y^2 \Phi^2 R + g \,  g_Y^4\Phi^4 \ . \,\,\, 
\eea
Now, the coefficients $f$ and $g$ are independent of the renormalization scheme but $d$ and $e$  are not. The finite Lagrangian counterterms required  by the renormalizability of the Yukawa interaction obtained in previous sections, 
\be \frac{\delta Z}{2} g^{\mu\nu}\nabla_\mu \Phi \nabla_\nu \Phi - \frac{\delta \xi_2}{2} R\Phi^2 - \frac{\delta \lambda_4}{4!} \Phi^4 \ , \ee
might alter the values of the coefficients $d$ and $e$, but not the coefficients $f$ and $g$. Note that, due to classical conformal invariance,  one should  consider only those counterterms having dimensionless coupling parameters.  Therefore, our results for the $f$ and $g$ coefficients are
\be \label{Tranomaly2}f = \frac{1}{3(4\pi)^2} N_f \ , \hspace{0.5cm} g= \frac{-1}{3(4\pi)^2} \left( \frac{3}{2}N_s- 6N_f \right) \ . \ee

Finally, to show explicitly that the above coefficients are independent of the renormalization scheme, we will compute them using the heat-kernel method given  in \cite{parker-toms}, by means of the one-loop effective action.

\subsection{Consistency with the heat-kernel results}

The conformal anomaly for a  field $\phi^j(x)$ obeying the second-order wave equation
\be \left[ \delta^i_j g^{\mu\nu}\nabla_\mu \nabla_\nu + Q^i_j(x) \right] \phi^j =0 \ , \ee
is given by 
\be C=\pm \frac{1}{(4\pi)^2} tr \ E_2(x) \ , \ee
where $E_2(x)$ is the second Seeley-DeWitt coefficient. The minus sign is for bosons and the plus sign is for fermions. These coefficients are local, scalar functions of $Q(x)$ and the curvature tensor. $E_2$ is given by
\be E_2= \left( -\frac{1}{30}\Box R + \frac{1}{72}R^2 -\frac{1}{180}R^{\mu\nu}R_{\mu\nu}+ \frac{1}{180}R^{\mu\nu\rho\sigma}R_{\mu\nu\rho\sigma} \right) I + \frac{1}{12}W^{\mu\nu}W_{\mu\nu} + \frac{1}{2}Q^2 -\frac{1}{6}RQ +\frac{1}{6}\Box Q \ , \ee
where $W_{\mu\nu}= [\nabla_\mu, \nabla_\nu]$. For a single massless scalar field with $\xi=1/6$ and an interaction of the form $h^2 \phi^2\Phi^2$ we have
\be Q= \frac{1}{6}R+ h^2\Phi^2 \ , \ee
and $W_{\mu\nu}=0$. For a spatially flat FLRW universe we get 
\be C= \frac{1}{2880\pi^2}\left[\Box R-\left
  (R^{\mu\nu}R_{\mu\nu}-\frac13R^2\right)\right] -\frac{h^2}{48\pi^2}(\Phi\Box \Phi +
  \nabla^{\mu}\Phi\nabla_{\mu}\Phi
  +\frac{3h^2}{2}\Phi^4) \ , \ee
 in full agreement with the result (\ref{ACs}) obtained using adiabatic regularization. 
  
   For a single massless Dirac field with  a Yukawa interaction we have [$(i\gamma^{\mu}\nabla_{\mu}-g_Y\Phi)\psi=0$]
\be Q= \left(\frac{1}{4}R+ g_Y^2\Phi^2\right) I + i  g_Y \gamma^{\mu}\nabla_{\mu} \Phi \ , \ee
and 
\be W_{\mu\nu}= -iR_{\mu\nu}^{\ \ \  \alpha\beta} \Sigma_{\alpha\beta}= -\frac{1}{8} R_{\mu\nu}^{\ \ \  \alpha\beta}[\gamma_{\alpha},\gamma_{\beta}]\ . \ee
Using the properties of the trace of products of gamma matrices, we get 
\be C= \frac{1}{2880\pi^2}\left[-11\left(R_{\alpha \beta}R^{\alpha \beta}-\frac13 R^2\right)+6\Box 
R\right]+\frac{g_Y^2}{8\pi^2}\left[- \frac{1}{3}\nabla^{\mu}\Phi\nabla_{\mu}\Phi+ \frac{2}{3}\Phi\Box \Phi + \frac{1}{6}  \Phi^2 R + g_Y^2\Phi^4
\right]. \  \ee
The above result reproduces the coefficients $f$ and $g$ obtained from adiabatic regularization. We note  that there is a mismatch in the  coefficients $d$ and $e$. These are, however, the coefficients that might depend on the renormalization scheme.

\section{Summary}\label{conclusions}
When a quantum field is coupled to a classical, nonadiabatic time-dependent background, it gets excited, and undergoes a regime of particle creation. In this case, new UV-divergent terms appear in the expectation values of its quadratic products, which must be appropriately removed to obtain a physical, finite quantity. In cosmological scenarios, adiabatic regularization provides an appropriate solution to this challenge: by means of an adiabatic expansion of the field modes, one can identify the covariant UV-divergent terms of the corresponding bilinear, and subtract them directly from the unrenormalized quantity. The background may be the expansion of the Universe itself, as in the case of inflation, or a classical homogeneous scalar field. The adiabatic scheme can be applied in both situations,  for both bosonic and fermionic species.

In this work, we have developed the adiabatic regularization method for spin-1/2 fields in an expanding universe, coupled to a classical background scalar field with a Yukawa interaction term. The results of this work are a natural generalization of the studies initiated in \cite{LNT, RNT}, and broaden significantly the range of applicability of the adiabatic method. We have computed the adiabatic expansion of the spin-1/2 field modes up to fourth adiabatic order, and used it to obtain expressions for the renormalized expectation values of the stress-energy tensor $\langle T_{\mu\nu} \rangle_{ren}$ and the bilinear $\langle \bar\psi\psi\rangle_{ren}$. These quantities are fundamental ingredients in the study of the semiclassical equations of fermionic matter interacting with a background field, as they codify the backreaction effects from the created matter on the metric/background fields.  Therefore, it is 
 essential to develop an efficient renormalization scheme to correctly quantify the effects of this backreaction. All expressions obtained are generic, depending only on the background scalar field and scale factor time-dependent functions. This constitutes probably the major advantage of the adiabatic renormalization scheme. We leave the method prepared to perform numerical computations in future investigations. 

Finally, we have tested the overall theoretical construction of the adiabatic scheme by justifying the method in terms of renormalization of coupling constants, as well as by computing the conformal anomaly. Our calculation of the conformal anomaly with the Yukawa interaction has been proved to be fully consistent with the generic results obtained via the one-loop effective action. Therefore,  by considering such a system, we have also improved our general understanding of quantum field theory in curved spacetimes.

\section*{Acknowledgments}

 F.T. thanks Daniel G. Figueroa and Juan Garcia-Bellido for useful discussions. This work is supported  by the Grants. No. FIS2014-57387-C3-1-P ,  No. FPA2015-68048-C3-3-P,  %No. CPANPHY-1205388, 
 No. MPNS of  COST Action No. CA15117, and the Severo Ochoa Programs SEV-2014-0398 and SEV-2012-0249.  A.d.R. is supported by  the FPU Ph.D. fellowship FPU13/04948, A. F. is supported by the Severo Ochoa Ph.D. fellowship SEV-2014-0398-16-1, and F.T. is supported by the Severo Ochoa Ph.D. fellowship SVP-2013-067697.

%F.T. thanks Daniel G. Figueroa and Juan Garcia-Bellido for useful discussions. This work is supported  by the  Research Projects of the Spanish MINECO FIS2014-57387-C3-1,  FPA2015-68048-C3-3-P,  the Consolider Program No. CPANPHY-1205388, the COST Action CA15117 (European Cooperation in Science and Technology), and the Centro de Excelencia Severo Ochoa Programs SEV-2014-0398 and SEV-2012-0249.  A.d.R. is supported by  the Spanish Ministry of Education Ph.D. fellowship FPU13/04948, A. F. is supported by the Severo Ochoa Ph.D. fellowship SEV-2014-0398-16-1, and F.T. is supported by the Severo Ochoa Ph.D. fellowship SVP-2013-067697.  

\appendix
\section{Scalar field with a Yukawa-type coupling}
\label{appendixA}

In this appendix we compute the conformal anomaly of a quantized real scalar field $\phi$, coupled to another background scalar $\Phi$ with a Yukawa-type interaction. This result will be used in Sec. \ref{sec:conformalanomaly}. The interaction term can be chosen of the form $g\Phi \phi^2$ or $h^2\Phi^2\phi^2$. Although the adiabatic regularization can be equally applied in both cases, we will focus on the latter case, since the coupling constant $h^2$ is dimensionless and the classical theory inherits the conformal invariance. Therefore, the action functional of the scalar matter field is given by
\be S_m= \int d^4x\sqrt{-g} \frac{1}{2}( g^{\mu\nu}\nabla_\mu \phi \nabla_\nu \phi -m^2\phi^2 -\xi R\phi^2 -h^2 \Phi^2\phi^2) \ . \ee 

As before, the scalar field lives in a spatially flat FLRW metric $ds^2=dt^2 -a^2(t)d\vec{x}^2$, and we assume that the external field is homogeneous $\Phi=\Phi(t)$. In this case, the equation of motion is
\be (\Box +m^2 +s^2(t)+\xi R)\phi=0 \label{motionscalar}\ , \ee
where we have introduced the notation $s(t)\equiv h \Phi(t)$, similar to the one used for the spin-$1/2$ field in the main text. The quantized field is expanded in Fourier modes as
\be \phi(x)= \frac{1}{\sqrt{2(2 \pi a^3)}}\int d^3 \vec{k} [A_{\vec{k}} f_{\vec{k}}(x)+A_{\vec{k}}^{\dagger}f_{\vec{k}}^{*}(x) ] \ , \label{phisolution} \ee
where $f_{\vec{k}}(x) = e^{i\vec{k}\vec{x}}h_k(t)$, and
 $A_{\vec{k}}^{\dagger}$ and $A_{\vec{k}}$ are the usual creation and annihilation operators. Substituting \eqref{phisolution} into \eqref{motionscalar} we find
\be \frac{d^2}{dt^2}h_k(t)+ \left[ \omega^2_k(t)+s^2(t)+\sigma(t) \right] h_k(t)=0 \label{equhk} \ , \ee
where $\sigma(t)=(6\xi- \frac{3}{4})(\frac{\dot{a}^2}{a^2})+(6\xi-\frac{3}{2})(\frac{\ddot{a}}{a})$, and $\omega_k(t)=\sqrt{\frac{k^2}{a(t)^2}+m^2}$.
The adiabatic expansion for the scalar field modes is based on the usual WKB ansatz
\be h_k(t)= \frac{1}{\sqrt{W_k}} e^{-i\int^t W_k (t') dt'} \ , \hspace{0.5cm}   W_k(t) = \omega_k + \omega^{(1)} + \omega^{(2)} + \cdots \ , \ee
which satisfies automatically the Wronskian condition $h_k\dot h_k^* - h_k^*\dot h_k = 2i$.  One can substitute  the ansatz into Eq. (\ref{equhk}), and solve order by order to obtain the different terms of the expansion. The function $W_k(t)$ obeys the differential equation
\be W_k^4= (\omega^2+s^2+\sigma)W_k^2+ \frac{3}{4} \dot{W}_k^2 - \frac{1}{2} \ddot{W}_k^2 W_k \ . \label{equWk2} \ee
Note that here, $s(t)\equiv h \Phi$ is assumed of adiabatic order one as in the fermionic case. One obtains systematically $\omega^{(odd)}=0$ for all terms of odd order in the expansion. At second adiabatic order one gets 
\begin{eqnarray}
  \omega^{(2)}&=&\frac{1}{2\omega}(s^2+\sigma)+\frac{3\dot{\omega}^2}{8\omega^3}-\frac{\ddot{\omega}}
  {4 \omega^2} \nonumber\\* &=&-\frac{m^2 \ddot{a}}{4 a\omega^3}+\frac{3 \xi \ddot{a}} {a \omega} 
  -\frac{\ddot{a}}{2 a \omega}+\frac{5 m^4 \dot{a}^2}{8 a^2 \omega^5} - \frac{m^2
   \dot{a}^2}{2 a^2 \omega^3}  + \frac{3 \xi \dot{a}^2}{a^2 \omega}
   -\frac{\dot{a}^2}{2 a^2\omega}+\frac{s^2}{2\omega} \ , 
   \end{eqnarray}
and at fourth adiabatic order, the result is
\begin{eqnarray}
  \omega^{(4)}&=&\frac{2(s^2+\sigma)\omega \omega^{(2)}+3/2 (\dot{\omega}^{(2)}\dot{\omega})
  -1/2\left[\ddot{\omega}^{(2)}\omega+\ddot{\omega}\omega^{(2)}\right]-s \omega^2 (\omega^{(2)})^2}
  {2 \omega^3}\nonumber \\* &=&
  -\frac{1105 \dot{a}^4 m^8}{128 a^4 \omega^{11}}+\frac{221 \ddot{a}\dot{a}^2
   m^6}{32 a^3 \omega^9}+\frac{221\dot{a}^4 m^6}{16 a^4 \omega^9}-\frac{7 \dot{a} a^{(3)} m^4}{8 a^2
    \omega^7}-\frac{25 s^2 \dot{a}^2 m^4}{16 a^2 
   \omega^7}-\frac{19\ddot{a}^2 m^4}{32 a^2 \omega^7}-\frac{75 \xi  \dot{a}^2 
  \ddot{a} m^4}{8 a^3 \omega^7}\nonumber\\* &-& \frac{111 \dot{a}^2 \ddot{a} m^4}{16 a^3 
   \omega ^7}-\frac{75 \xi \dot{a}^4 m^4}{8 a^4 \omega^7}-\frac{69 \dot{a}^4 m^4}
   {16a^4 \omega^7}+\frac{9 \ddot{a}^2 \xi  m^2}{4 a^2 \omega^5}+\frac{15\dot{a} a^{(3)} \xi  m^2}
   {4 a^2 \omega^5}+\frac{18\ddot{a}\dot{a}^2 \xi  
   m^2}{a^3 \omega^5}+\frac{9 \dot{a}^4 \xi  m^2}{2 a^4 \omega^5}\nonumber\\* &+&\frac{5 s \dot{a}
    \dot{s} m^2}{4 a \omega^5}+\frac{3 \ddot{a} s^2 m^2}{8 a\omega
    ^5}+\frac{a^{(4)} m^2}{16 a \omega^5}+\frac{2 s^2 \dot{a}^2 m^2}{a^2 
    \omega^5}+\frac{\ddot{a}^2 m^2}{16 a^2 \omega^5}+\frac{\dot{a} a^{(3)} 
    m^2}{16 a^2 \omega^5}-\frac{15 \dot{a}^2 \ddot{a} m^2}{16 a^3 \omega^5}
    -\frac{\dot{a}^4 m^2}{4 a^4 \omega^5}\nonumber\\* &+& \frac{3\ddot{a} \dot{a}^2 \xi }
    {4 a^3 \omega^3}+\frac{3 \dot{a}^4 \xi }{2 a^4 \omega^3}+\frac{a^{(4)}}{8 a 
    \omega^3}-\frac{\dot{s}^2}{4 \omega^3}-\frac{s\ddot{s}}{4 \omega^3}-
    \frac{s^4}{8 \omega^3}-\frac{3 \xi  s^2 \ddot{a}}{2 a \omega^3}-\frac{5 s
   \dot{a}\dot{s}}{4 a \omega^3}-\frac{3 \xi a^{(4)}}{4 a \omega^3}
    -\frac{3 \xi  s^2 \dot{a}^2}{2 a^2 \omega^3}\nonumber\\* &-& \frac{9 \xi ^2 \ddot{a}^2}{2 a^2 
    \omega^3}-\frac{s^2 \dot{a}^2}{4 a^2 \omega^3}-\frac{3 \xi  \ddot{a}^2}{4 
    a^2 \omega^3}+\frac{\ddot{a}^2}{4 a^2 \omega^3}-\frac{15 \xi \dot{a}
    a^{(3)}}{4a^2 \omega^3}+\frac{5 \dot{a} a^{(3)} }{8 a^2 \omega^3}-\frac{9 \xi ^2 \dot{a}^2\ddot{a}}
    {a^3 \omega^3}+\frac{\ddot{a} \dot{a}^2}{8 a^3 \omega^3} \nonumber\\* &-& \frac{9 \xi ^2 \dot{a}^4}
    {2 a^4 \omega^3}-\frac{\dot{a}^4}{8 a^4 \omega^3} \ .
\end{eqnarray}
Expressions for the subtraction terms in conformal time have been obtained in \cite{Molina-Paris-Anderson-Ramsey}. Here we will briefly sketch the  renormalization counterterms associated to the UV divergences of the stress-energy tensor and the variance $\langle \phi^2\rangle$. We follow a strategy similar to the one used in Sec. {\ref{sectioncounterterms}. The Lagrangian density with the required renormalization counterterms  is 
\bea \mathcal L &=& \mathcal L_m + {\sqrt{-g}}\left[ \frac{1}{2}g^{\mu\nu}\nabla_\mu \Phi \nabla_\nu \Phi - \frac{m^2}{2}\Phi^2 - \frac{\lambda}{4!}\Phi^4-\frac{1}{2}\xi_2R \Phi^2 - \frac{1}{8\pi G}\Lambda + \frac{1}{16\pi G}R + \alpha R^2\right] \nonumber \\
&+& {\sqrt{-g}}\left[ \frac{1}{2}\delta Zg^{\mu\nu}\nabla_\mu \Phi \nabla_\nu \Phi -  \frac{\delta m^2 }{2}\Phi^2 -  \frac{\delta \lambda }{4!}\Phi^4 -\frac{1}{2}\delta {\xi_2}R \Phi^2 - \frac{1}{8\pi}\delta {\Lambda} + \frac{1}{16\pi }\delta {G^{-1}}R + \delta \alpha R^2\right]\ . \ \ \ \ \ \ \ \ \eea 
One can check that the above counterterms are enough to absorb all the UV divergences that emerge in the quantization of the scalar field.
We note that, due to the symmetry $\Phi \to -\Phi$ of the matter Lagrangian, counterterms of the form $R \Phi$, $\Phi$, $\Phi^3$ are absent. However, a higher-derivative term of the form $R^2$ is now necessary, which did not appear for the Dirac field in a FLRW spacetime.  

We assume the conformal coupling to the curvature $\xi=1/6$.
For a massive field we have
\be g^{\mu\nu} T_{\mu\nu} - h^2\Phi^2 \phi^2 =m^2\phi^2\ . \ee
Classical conformal invariance is obtained when $m^2=0$. In adiabatic regularization the conformal anomaly is computed by taking the massless limit
\be C_s= g^{\mu\nu} \langle T_{\mu\nu}\rangle  -h^2 \Phi^2 \langle \phi^2 \rangle = -\lim_{m^2\to 0} m^2 \langle \phi^2\rangle^{(4)}\nonumber =-\lim_{m^2 \to 0} m^2(4\pi 
  a^3)^{-1}\int^{\infty}_{0}dk k^2 (W_k^{-1}(t))^{(4)} \ ,\ee
where  $(W_k^{-1}(t))^{(4)}= \omega^{-3}(\omega^{(2)})^2-\omega^{-2}\omega^{(4)}$ is the fourth-order term in the adiabatic expansion of $W_k^{-1}$. Note that here, $\langle \phi^2 \rangle^{(4)}$ is evaluated including fourth-order adiabatic subtractions. This is different to the physical vacuum expectation value $\langle \phi^2 \rangle_{ren}$, which has to be evaluated with subtractions only up to second order. This is why only the purely fourth-order adiabatic  piece contributes to the anomaly.  The explicit expression of $(W_k^{-1})^{(4)}$ for arbitrary $\xi$ is
\bea
(W_k^{-1})^{(4)} (t)=&+&\frac{1155 \dot{a}^4 m^8}{128 a^4 \omega ^{13}}-\frac{231 \dot{a}^2\ddot{a} 
m^6}{32 a^3 \omega ^{11}}-\frac{231 \dot{a}^4 m^6}{16 a^4 \omega ^{11}}  +\frac{105 \xi  \dot{a}^4 m^4}
{8 a^4 \omega ^9}+\frac{63 \dot{a}^4 m^4}{16 a^4 \omega ^9}+\frac{35 s^2 \dot{a}^2 m^4}{16 a^2 \omega
 ^9}\nonumber \\* 
 &+&\frac{105\ddot{a} \xi  \dot{a}^2 m^4}{8 a^3 \omega ^9}+\frac{105 \ddot{a} \dot{a}^2 
 m^4}{16 a^3 \omega ^9}+\frac{7 a^{(3)} \dot{a} m^4}{8 a^2 \omega ^9}+\frac{21 \ddot{a}^2 m^4}{32 a^2 
 \omega ^9}+\frac{3 \dot{a}^4 m^2}{4 a^4 \omega ^7}+\frac{27 \ddot{a}
 \dot{a}^2 m^2}{16 a^3 \omega ^7}\nonumber \\* &-&\frac{5 \dot{a} s \dot{s} m^2}{4 a \omega ^7}-\frac{5 s^2\ddot{a}
  m^2}{8 a \omega ^7}-\frac{a^{(4)} m^2}{16 a \omega ^7}-\frac{5 \dot{a}^2 s^2 m^2}
 {2 a^2 \omega ^7}-\frac{15 \xi  \ddot{a}^2 m^2}{4 a^2 \omega ^7}-\frac{15 \dot{a} \xi  a^{(3)}
  m^2}{4 a^2 \omega ^7}+ \frac{3\ddot{a}^2 m^2}{16 a^2 \omega ^7}\nonumber \\&-&\frac{\dot{a} a^{(3)}
   m^2}{16 a^2 \omega ^7}-\frac{45 \dot{a}^2 \xi  \ddot{a} m^2}{2 a^3 \omega ^7}-\frac{15 
   \dot{a}^4 \xi  m^2}{2 a^4 \omega ^7}+\frac{27 \xi ^2 \dot{a}^4}{2 a^4 \omega ^5}+\frac{3 \dot{a}^4}{8 a^4
    \omega ^5}+\frac{9 s^2 \xi  \dot{a}^2}{2 a^2 \omega ^5}+\frac{27 \ddot{a} \xi ^2 \dot{a}^2}{a^3 
    \omega ^5}\nonumber \\*
    &+&\frac{3 \ddot{a} \dot{a}^2}{8 a^3 \omega ^5}+\frac{\dot{s}^2}{4 \omega ^5}+
    \frac{15 a^{(3)} \xi  \dot{a}}{4 a^2 \omega ^5}+\frac{5 s \dot{a} \dot{s}}{4 a \omega ^5}+
    \frac{s \ddot{s}}{4 \omega ^5}+\frac{3 s^4}{8 \omega ^5}-\frac{s^2 \ddot{a}}{2 a 
    \omega ^5}+\frac{9 s^2\ddot{a} \xi }{2 a \omega ^5} +\frac{3 a^{(4)} \xi }{4 a
     \omega ^5}\nonumber
    \\* &-&\frac{a^{(4)}}{8 a \omega ^5}+\frac{27 \ddot{a}^2 \xi ^2}{2 a^2 \omega ^5}-
     \frac{\dot{a}^2 s^2}{4 a^2 \omega ^5}-\frac{9 \xi \ddot{a}^2}{4 a^2 \omega ^5}-\frac{5 
     \dot{a} a^{(3)}}{8 a^2 \omega ^5}-\frac{27 \dot{a}^2 \xi  \ddot{a}}{4 a^3 \omega ^5}
     -\frac{9 \dot{a}^4 \xi }{2 a^4 \omega ^5} \ .
     \eea
     
The integral in comoving momenta is finite and independent of the mass. Assuming now $\xi=1/6$, the result is 
\be C_s= \frac{a^{(4)}}{480 \pi ^2 a}+\frac{\ddot{a}^2}{480 \pi ^2 a^2}-\frac{s\dot{a} \dot{s}}
{16 \pi ^2 a}  +\frac{a^{(3)} \dot{a}}{160 \pi ^2 a^2}-\frac{\dot{a}^2 \ddot{a}}
{160 \pi ^2 a^3}  -\frac{s\ddot{s}}{48 \pi ^2}
-\frac{\dot{s}^2}{48 \pi ^2}-\frac{s^4}{32 \pi ^2}
 \ .  \ee
We can rewrite the expression in terms of covariant scalar terms as
\be
  C_s=\frac{1}{2880\pi^2}\left\{\Box R-\left
  (R^{\mu\nu}R_{\mu\nu}-\frac13R^2\right)\right\}-\frac{h^2}{48\pi^2}(\Phi\Box \Phi +
  \nabla^{\mu}\Phi\nabla_{\mu}\Phi
  +\frac{3h^2}{2}\Phi^4) \ , \ee
which is the result given in Eq. (\ref{ACs}).

\section{A simple example}
\label{appendixB}
In this appendix we consider a simple mathematical example to illustrate how the adiabatic method works. We compute the bilinear $\langle \bar{\psi} \psi \rangle_{\rm ren}$ of a Dirac field, coupled to a background scalar field evolving in Minkowski spacetime [$a(t) = 1$] as 
\be s(t) = g_Y\Phi(t)= \mu/t \ .\label{eq:smut}\ee 
For convenience, we have absorbed the Yukawa coupling $g_Y$ in the dimensionless constant $\mu$. To avoid the mathematical instability at $t \rightarrow 0$, we will only consider times in the range $-\infty < t < 0$. This model has three convenient aspects which simplify significantly the analysis. First of all, the field mode equations (\ref{ferm-hk2b}) have an analytical solution in terms of the well-known Whitakker functions, so we do not have to solve the equation numerically. Second, at time $t \rightarrow - \infty$ we have $s, \dot s \cdots \rightarrow 0$, so that the system is adiabatic initially, and there is no ambiguity when imposing initial conditions to the field modes. And third, as we shall see, the system behaves in such a way that, as long as we are well before the instability,  $\langle \bar{\psi} \psi \rangle_{\rm ren}$ can be approximated by the fourth order in its adiabatic expansion, giving a final renormalized bilinear that can be easily integrated.

It is useful to define a new dimensionless time $z\equiv m t$ and momenta $\kappa \equiv k / m$. The field equations (\ref{ferm-hk2b}) for $h_k^{I}$ and $h_k^{II}$ in terms of these variables become
\be  h_k^{II} = \frac{i}{\kappa} \left[ \frac{\partial h_k^I}{\partial z} + i \left( 1 + \frac{\mu}{z} \right) h_k^I \right] \ , \hspace{0.5cm} h_k^{I} =  \frac{i }{\kappa} \left[ \frac{\partial h_k^{II}}{\partial z} - i \left(1 + \frac{\mu}{z}\right)h_k^{II} \right] \label{ferm-hk3} \ , \ee
and from these, we obtain the second-order uncoupled equations
 \be \frac{d^2 h_k^{I}}{d z^2} + \left( 1 +\kappa^2 +\frac{2 \mu}{z}+ \frac{\mu(\mu - i)}{z^2} \right) h_k^{I}=0 \ , \hspace{0.5cm}
  \frac{d^2 h_k^{II}}{dz^2} + \left( 1 +\kappa^2 + \frac{2\mu}{z} + \frac{\mu(\mu + i)}{z^2} \right) h_k^{II}=0 \ . \ee 
Let us also define a dimensionless frequency $\omega_{\kappa} \equiv \sqrt{\kappa^2 + 1}$, so that $\omega = \sqrt{ k^2 + m^2} = m \omega_{\kappa}$. The general solution for $h^{I}_k (t)$ is a linear combination of the first and second kind Whittaker functions $M_{\alpha,\lambda_1}\left(2 i\omega_{\kappa} t\right)$ and $W_{\alpha,\lambda_1}\left(2 i \omega_{\kappa} t\right)$, where $\alpha \equiv \frac{-i \mu }{\sqrt{\kappa^2+1}}$ and  $\lambda_1\equiv -\frac{1}{2} i (2 \mu -i)$. The solution for $h^{II}_k (t)$   is similar, with the change $\lambda_1 \to \lambda_2 \equiv-\frac{1}{2} i(2 \mu +i)$, so we have
\bea h^{I}_k&=&A_k^I M_{\alpha,\lambda_1}\left(2 i \omega_{\kappa} z\right)+B_k^{I} W_{\alpha,\lambda_1}\left(2 i \omega_{\kappa} z \right) \ , \nonumber \\*
  h^{II}_k&=&A_k^{II} M_{\alpha,\lambda_2}\left(2 i \omega_{\kappa} z\right)+B_k^{II} W_{\alpha,\lambda_2}\left(2 i\omega_{\kappa} z \right) \ . \eea
Note that $h_k^I$ and $h_k^{II}$ must obey the constraint (\ref{ferm-wronsk2}), so  there is only 1 degree of freedom in the fermion solution, which is determined when imposing the initial conditions. To fix the constants in the linear combinations, we impose the adiabatic behavior (\ref{ferm-minkowski}) at $z\to -\infty$, getting
\be
A_k^{I} = A_k^{II}=0 \ , \hspace{0.5cm} B_k^{I} = \sqrt{\frac{\omega_{\kappa} +1}{2\omega_{\kappa}}}e^{\frac{\mu \pi}{2\omega_{\kappa}}} \ , \hspace{0.5cm} B_k^{II} = \sqrt{\frac{\omega_{\kappa} -1}{2\omega_{\kappa} }}e^{\frac{\mu \pi}{2\omega_{\kappa}}} \ . \ee
The final solution is then
\be h^{I}_k=\sqrt{\frac{\omega_{\kappa} + 1}{2\omega_{\kappa} }} e^{\frac{\mu \pi}{2\omega_{\kappa}}}W_{\alpha,\lambda_1}\left(2 i \omega_{\kappa} z\right) \ , \hspace{0.5cm}
h^{II}_k=\sqrt{\frac{\omega_{\kappa}-1}{2\omega_{\kappa}}} e^{\frac{\mu \pi}{2\omega_{\kappa}}} W_{\alpha,\lambda_2}\left(2 i\omega_{\kappa} z \right) \ .\ee

\begin{figure}
\begin{center}
\begin{tabular}{c}
\includegraphics[width=10.3cm]{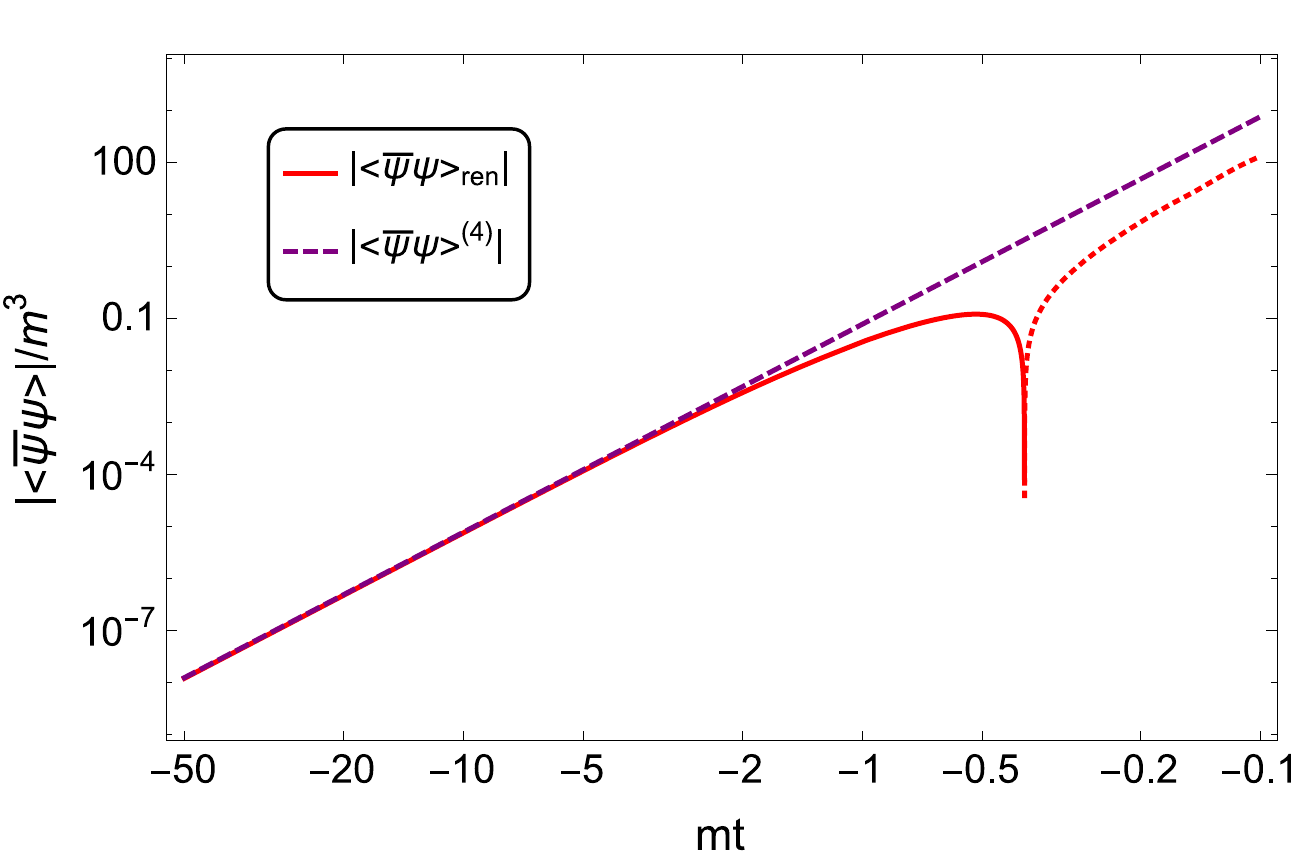}
\end{tabular}
\end{center}
\caption{The red line shows $\frac{1}{m^3} | \langle \bar{\psi} \psi \rangle_{\rm ren} | $ as a function of time for $\mu=1$ given by Eq. (\ref{eq:example-ren}). For $mt \lesssim - 0.4$ we have $\langle \bar{\psi} \psi \rangle_{\rm ren} < 0$ (red continuous line), while for $mt \gtrsim - 0.4$  we have $\langle \bar{\psi} \psi \rangle_{\rm ren}  > 0$ (red dotted line). The purple dashed line shows the corresponding approximation at fourth adiabatic order given in Eq. (\ref{eq:example-4thorder}).}
\label{psi}
\end{figure}

The renormalized expectation value $\langle \bar{\psi}\psi \rangle_{ren}$ is given, from (\ref{eq:ren-variance}), by 
\bea \langle \bar{\psi} \psi\rangle_{ren} = \frac{-m^3}{\pi^2}\int_0^{\infty} d \kappa \kappa^2 && \left( |h_k^{I}|^2-|h_k^{II}|^2 - \frac{1}{\omega_{\kappa}}-\frac{\mu}{\omega_{\kappa} z} +\frac{\mu}{\omega_{\kappa}^3 z}+\frac{3\mu^2}{2\omega_{\kappa}^3 z^2} - \frac{3 \mu^2}{2\omega_{\kappa}^5 z^2} \right. \nonumber \\* 
&&+ \left. \frac{\mu+\mu^3}{2\omega_{\kappa}^3 z^3} - \frac{\mu+6 \mu^3}{2\omega_{\kappa}^5 z^3} +\frac{5 \mu^3}{2\omega_{\kappa}^7z^3} \right) \label{eq:example-ren} \ , \eea  
where we have that the adiabatic contributions of order $n$ go as $\langle \bar{\psi} \psi \rangle^{(n)} \propto c_n(\mu,\kappa) z^{-n}$, with $c_n$ time-independent functions of $\mu$ and $\kappa$. The above integral is finite, as one can easily check from the asymptotic expansion of the Whittaker function $W_{\alpha, \lambda}(x)$. 

We can compute analytically the leading  term at $z \to -\infty$ by performing the adiabatic expansion of $\langle \bar{\psi}\psi \rangle$ up to fourth order, and subtracting from it the zeroth, first, second, and third orders. Therefore, the leading behavior at very early times is
\bea \langle \bar{\psi} \psi\rangle_{ren} \sim  \langle \bar{\psi} \psi\rangle^{(4)}&\equiv&-\frac{1}{\pi^2a^3} 
 \int^{\infty}_{0}dk k^2 \left(\left(|h_k^{I}|^2\right)^{(4)}-\left(|h_k^{II}|^2\right)^{(4)}\right)\nonumber \\
&=&-\frac{1}{\pi^2a^3} 
 \int^{\infty}_{0}dk k^2 \left( \frac{(\omega -m)}{2 \omega}[ G^{(4)} + G^{(4)*} + G^{(1)}G^{(3)*} + G^{(1)*}G^{(3)} + |G^{(2)}|^2] \right. \nonumber \\*
 && \left. -\frac{(\omega +m)}{2 \omega}[ F^{(4)} + F^{(4)*} + F^{(1)}F^{(3)*} + F^{(1)*}F^{(3)} + |F^{(2)}|^2] \right) \ . \eea    
Computing  the integral, we finally get
\bea \langle \bar{\psi} \psi\rangle^{(4)} &=&  -\frac{a^{(4)}}{80 \pi ^2 a m}+\frac{\dot{a}^2 \ddot{a}}{60 \pi ^2 a^3 m}-\frac{\dot{a}^2 s^2}{8 \pi ^2 a^2 m}-\frac{\ddot{a}^2}{80 \pi ^2 a^2
   m}-\frac{3 \dot{a} a^{(3)}}{80 \pi ^2 a^2 m}-\frac{s^2 \ddot{a}}{8 \pi ^2 a m}\nonumber \\ &&-\frac{3 \dot{a} \dot{s} s}{4 \pi ^2 a m}-\frac{s^4}{8 \pi ^2
   m}-\frac{s \ddot{s}}{4 \pi ^2 m}-\frac{\dot{s}^2}{8 \pi ^2 m} \ . \eea
Substituting (\ref{eq:smut}) in this expression, and setting $a=1$, we finally obtain
\be \langle \bar{\psi} \psi\rangle^{(4)} = -\frac{m^3 \mu ^2(\mu^2+5)}{8 \pi ^2 z^4} \ , \label{eq:example-4thorder} \ee
where we have written the solution in terms of $z$. In Fig. \ref{psi} we show $\frac{1}{m^3} | \langle \bar{\psi} \psi \rangle |_{ren}$ as a function of time, comparing the exact result (\ref{eq:example-ren}) with the approximation (\ref{eq:example-4thorder}). At very early times $z \to -\infty$ we have, as expected, $ \langle \bar{\psi} \psi\rangle_{ren} \sim 0$. We observe that the approximation holds quite well, except when the instability is approached.

\section{Adiabatic expansion}\label{appendixC}

In this appendix, we provide the terms of the adiabatic expansion of the spin-1/2 field modes up to fourth order. Although the first- and second-order terms have already been written in Section \ref{adiabaticexpansion}, we copy them here for convenience. As introduced in Eqs. (\ref{ferm-ansatz}) and (\ref{ferm-expansions}), the adiabatic expansion takes the form
  \bea h_k^{I} (t) &=& \sqrt{\frac{\omega + m}{2 \omega}} e^{-i \int^t ( \omega + \omega^{(1)} + \omega^{(2)} + \omega^{(3)} + \omega^{(4)} + \dots) dt'} (1 + F^{(1)} + F^{(2)} + F^{(3)} + F^{(4)} + \dots) \ , \nonumber \\
   h_k^{II} (t) &=& \sqrt{\frac{\omega - m}{2 \omega}} e^{-i \int^t ( \omega + \omega^{(1)} + \omega^{(2)} + \omega^{(3)} + \omega^{(4)} + \dots) dt'} (1 + G^{(1)} + G^{(2)} + G^{(3)} + G^{(4)} + \dots) \ . \label{ferm-ansatz-app} \eea
The terms $G^{(n)}$ can be obtained from $F^{(n)}$ with the relation $G^{(n)} (m, s) = F^{(n)} (-m,-s)$, so we do not explicitly write them here. We denote by $f_x^{(n)}$ and $f_y^{(n)}$ to the real and imaginary parts of $F^{(n)}$ respectively, so that $F^{(n)} = f_x^{(n)} + i f_y^{(n)}$. 

The first-order terms are
\bea f^{(1)}_x &=& \frac{s}{2 \omega} - \frac{m s}{2 \omega^2}  \ , \\ 
f^{(1)}_y &=& - \frac{m \dot{a}}{4 \omega^2 a}  \ , \\
\omega^{(1)} &=& \frac{m s}{\omega} \ .\eea

The second-order terms are
\bea f^{(2)}_x &=& \frac{m^2 \ddot{a}}{8 a \omega^4}-\frac{m \ddot{a}}{8 a \omega^3}-\frac{5 m^4 \dot{a}^2}{16 a^2 \omega^6}+\frac{5 m^3 \dot{a}^2}{16 a^2 \omega^5}+\frac{3 m^2 \dot{a}^2}{32 a^2 \omega^4}-\frac{m \dot{a}^2}{8 a^2 \omega^3}+\frac{5 m^2 s^2}{8 \omega^4}-\frac{m s^2}{2 \omega^3}-\frac{s^2}{8 \omega^2} \ , \\
f^{(2)}_y &=& \frac{5 m^2 s \dot{a}}{8 a \omega^4} - \frac{s \dot{a}}{4 a \omega^2} - \frac{\dot{s}}{4 \omega^2} \ , \\
\omega^{(2)} &=& -\frac{m^2 s^2}{2 \omega^3} + \frac{s^2}{2 \omega} + \frac{5 m^4 \dot{a}^2}{8 a^2 \omega^5} - \frac{3 m^2 \dot{a}^2}{8 a^2 \omega^3} - \frac{m^2 \ddot{a}}{4 a \omega^3} \ .\eea

The third-order terms are
\bea f^{(3)}_x &=& - \frac{15 m^3 s^3}{16 \omega^6} + \frac{11 m^2 s^3}{16 \omega^5} + \frac{ 7 m s^3}{16 \omega^4} - \frac{3 s^3}{16 \omega^3} + \frac{65 m^5 s \dot{a}^2}{32 a^2 \omega^8} - \frac{15 m^4 s \dot{a}^2}{8 a^2 \omega^7} - \frac{97 m^3 s \dot{a}^2}{64 a^2 \omega^6} + \frac{93 m^2 s \dot{a}^2}{64 a^2 \omega^5} \nonumber  \\*
  &&+ \frac{m s \dot{a}^2}{8 a^2 \omega^4} - \frac{s \dot{a}^2}{8 a^2 \omega^3} - \frac{5 m^3 \dot{a} \dot{s}}{8 a \omega^6}  + \frac{5 m^2 \dot{a} \dot{s}}{8 a \omega^5} + \frac{5 m \dot{a} \dot{s}}{16 a \omega^4} - \frac{3 \dot{a} \dot{s}}{8 a \omega^3} - \frac{9 m^3 s \ddot{a}}{16 a \omega^6} + \frac{m^2 s \ddot{a}}{2 a \omega^5} + \frac{3 m s \ddot{a}}{16 a \omega^4} \nonumber \\*
  &&- \frac{s\ddot{a}}{8 a \omega^3}  + \frac{m \ddot{s}}{8 \omega^4} - \frac{\ddot{s}}{8 \omega^3} \ , \\
f^{(3)}_y &=& - \frac{45 m^3 s^2 \dot{a}}{32 a \omega^6} +\frac{31 m s^2 \dot{a}}{32 a \omega^4} + \frac{65 m^5 \dot{a}^3}{64 a^3 \omega^8} - \frac{97 m^3 \dot{a}^3}{128 a^3 \omega^6} + \frac{m \dot{a}^3}{16 a^3 \omega^4} + \frac{5 m s \dot{s}}{8 \omega^4} - \frac{19 m^3 \dot{a} \ddot{a}}{32 a^2 \omega^6} + \frac{m \dot{a} \ddot{a}}{4 a^2 \omega^4} \nonumber \\*
 && + \frac{m a^{(3)}}{16 a \omega^4}  \ , \\
\omega^{(3)} &=& \frac{m^3 s^3}{2 \omega^5} - \frac{m s^3}{2 \omega^3} - \frac{25 m^5 s \dot{a}^2}{8 a^2 \omega^7} + \frac{13 m^3 s \dot{a}^2}{4 a^2 \omega^5} - \frac{m s \dot{a}^2}{2 a^2 \omega^3} + \frac{5 m^3 \dot{a} \dot{s}}{4 a \omega^5} - \frac{7 m \dot{a} \dot{s}}{8 a \omega^3} + \frac{3 m^3 s \ddot{a}}{4 a \omega^5} - \frac{3 m s \ddot{a}}{8 a \omega^3} - \frac{m \ddot{s}}{4 \omega^3} \nonumber \ . \\  \eea

Finally, the fourth-order terms are
\bea f^{(4)}_x&=&\frac{2285 \dot{a}^4 m^8}{512 a^4 \omega ^{12}}-\frac{565 \dot{a}^4 m^7}{128 a^4 \omega
   ^{11}}-\frac{1263 \dot{a}^4 m^6}{256 a^4 \omega ^{10}}-\frac{1105 s^2 \dot{a}^2
   m^6}{128 a^2 \omega ^{10}}-\frac{457 \dot{a}^2 \ddot{a} m^6}{128 a^3 \omega
   ^{10}}+\frac{2611 \dot{a}^4 m^5}{512 a^4 \omega ^9}+\frac{965 s^2 \dot{a}^2 m^5}{128
   a^2 \omega ^9}\nonumber \\ && +\frac{113 \dot{a}^2 \ddot{a} m^5}{32 a^3 \omega ^9}+\frac{2371
   \dot{a}^4 m^4}{2048 a^4 \omega ^8}+\frac{2441 s^2 \dot{a}^2 m^4}{256 a^2 \omega
   ^8}+\frac{41 \ddot{a}^2 m^4}{128 a^2 \omega ^8}+\frac{65 s \dot{a} \dot{s} m^4}{16 a
   \omega ^8}+\frac{725 \dot{a}^2 \ddot{a} m^4}{256 a^3 \omega ^8}+\frac{117 s^2
   \ddot{a} m^4}{64 a \omega ^8}\nonumber \\ &&+\frac{7 \dot{a} a^{(3)} m^4}{16 a^2 \omega
   ^8}+\frac{195 s^4 m^4}{128 \omega ^8}-\frac{333 \dot{a}^4 m^3}{256 a^4 \omega
   ^7}-\frac{1049 s^2 \dot{a}^2 m^3}{128 a^2 \omega ^7}-\frac{5 \ddot{a}^2 m^3}{16 a^2
   \omega ^7}-\frac{15 s \dot{a} \dot{s} m^3}{4 a \omega ^7}-\frac{749 \dot{a}^2
   \ddot{a} m^3}{256 a^3 \omega ^7}\nonumber \\ &&-\frac{97 s^2 \ddot{a} m^3}{64 a \omega ^7}-\frac{7
   \dot{a} a^{(3)} m^3}{16 a^2 \omega ^7}-\frac{17 s^4 m^3}{16 \omega ^7}-\frac{3
   \dot{a}^4 m^2}{128 a^4 \omega ^6}-\frac{561 s^2 \dot{a}^2 m^2}{256 a^2 \omega
   ^6}-\frac{5 \dot{s}^2 m^2}{16 \omega ^6}-\frac{17 \ddot{a}^2 m^2}{128 a^2 \omega
   ^6}\nonumber \\ && -\frac{95 s \dot{a} \dot{s} m^2}{32 a \omega ^6}-\frac{19 \dot{a}^2 \ddot{a}
   m^2}{64 a^3 \omega ^6}-\frac{73 s^2 \ddot{a} m^2}{64 a \omega ^6}-\frac{9 s \ddot{s}
   m^2}{16 \omega ^6}-\frac{13 \dot{a} a^{(3)} m^2}{64 a^2 \omega
   ^6}-\frac{a^{(4)} m^2}{32 a \omega ^6}-\frac{71 s^4 m^2}{64 \omega
   ^6}\nonumber \\ &&+\frac{\dot{a}^4 m}{32 a^4 \omega ^5}+\frac{111 s^2 \dot{a}^2 m}{64 a^2 \omega
   ^5}+\frac{5 \dot{s}^2 m}{16 \omega ^5}+\frac{\ddot{a}^2 m}{8 a^2 \omega ^5}+\frac{89
   s \dot{a} \dot{s} m}{32 a \omega ^5}+\frac{11 \dot{a}^2 \ddot{a} m}{32 a^3 \omega
   ^5}+\frac{49 s^2 \ddot{a} m}{64 a \omega ^5}+\frac{s \ddot{s} m}{2 \omega
   ^5}\nonumber \\ &&+\frac{7 \dot{a} a^{(3)} m}{32 a^2 \omega ^5}+\frac{a^{(4)} m}{32 a \omega
   ^5}+\frac{9 s^4 m}{16 \omega ^5}+\frac{s^2 \dot{a}^2}{32 a^2 \omega
   ^4}-\frac{\dot{s}^2}{32 \omega ^4}+\frac{s \dot{a} \dot{s}}{8 a \omega ^4}+\frac{s^2
   \ddot{a}}{16 a \omega ^4}+\frac{s \ddot{s}}{16 \omega ^4}+\frac{11 s^4}{128 \omega
   ^4} \ , \\ \nonumber \\
    f^{(4)}_y &=& \frac{195 m^4 s^3 \dot{a}}{64 a \omega^8} - \frac{187 m^2 s^3 \dot{a}}{64 a \omega^6} + \frac{11 s^3 \dot{a}}{32 a \omega^4} - \frac{1105 m^6 s \dot{a}^3}{128 a^3 \omega^{10}} + \frac{2571 m^4 s \dot{a}^3}{256 a^3 \omega^8} - \frac{329 m^2 s \dot{a}^3}{128 a^3 \omega^6} + \frac{s \dot{a}^3}{16 a^3 \omega^4}\nonumber \\*
&& - \frac{45 m^2 s^2 \dot{s}}{32 \omega^6} + \frac{11 s^2 \dot{s}}{32 \omega^4}   + \frac{195 m^4 \dot{a}^2 \dot{s}}{64 a^2 \omega^8} - \frac{367 m^2 \dot{a}^2 \dot{s}}{128 a^2 \omega^6} + \frac{7 \dot{a}^2 \dot{s}}{16 a^2 \omega^4} + \frac{247 m^4 s \dot{a} \ddot{a}}{64 a^2 \omega^8} - \frac{187 m^2 s \dot{a} \ddot{a}}{64 a^2 \omega^6} + \frac{s \dot{a} \ddot{a}}{4 a^2 \omega^4}\nonumber \\* 
  && - \frac{19 m^2 \dot{s} \ddot{a}}{32 a \omega^6} + \frac{\dot{s} \ddot{a}}{4 a \omega^4}  - \frac{19 m^2 \dot{a} \ddot{s}}{32 a \omega^6} + \frac{3 \dot{a} \ddot{s}}{8 a \omega^4}  - \frac{9 m^2 s a^{(3)}}{32 a \omega^6} + \frac{s a^{(3)}}{16 a \omega^4} + \frac{s^{(3)}}{16 \omega^4} \ , \\ \nonumber \\
 \omega^{(4)} = &-& \frac{5 m^4 s^4}{8 \omega^7} + \frac{3 m^2 s^4}{4 \omega^5} - \frac{s^4}{8 \omega^3} + \frac{175 m^6 s^2 \dot{a}^2}{16 a^2 \omega^9} - \frac{245 m^4 s^2 \dot{a}^2}{16 a^2 \omega^7} + \frac{79 m^2 s^2 \dot{a}^2}{16 a^2 \omega^5} - \frac{s^2 \dot{a}^2}{8 a^2 \omega^3} - \frac{1105 m^8 \dot{a}^4}{128 a^4 \omega^{11}} \nonumber \\*
 &&+ \frac{337 m^6 \dot{a}^4}{32 a^4 \omega^9} - \frac{377 m^4 \dot{a}^4}{128 a^4 \omega^7} + \frac{3 m^2 \dot{a}^4}{32 a^4 \omega^5} - \frac{25 m^4 s \dot{a} \dot{s}}{4 a \omega^7} + \frac{23 m^2 s \dot{a} \dot{s}}{4 a \omega^5} - \frac{3 s \dot{a} \dot{s}}{8 a \omega^3} + \frac{5 m^2 \dot{s}^2}{8 \omega^5} - \frac{15 m^4 s^2 \ddot{a}}{8 a \omega^7}  \nonumber \\*
&&+ \frac{25 m^2 s^2 \ddot{a}}{16 a \omega^5} - \frac{s^2 \ddot{a}}{8 a \omega^3} + \frac{221 m^6 \dot{a}^2 \ddot{a}}{32 a^3 \omega^9}  - \frac{389 m^4 \dot{a}^2 \ddot{a}}{64 a^3 \omega^7} + \frac{13 m^2 \dot{a}^2 \ddot{a}}{16 a^3 \omega^5} - \frac{19 m^4 \ddot{a}^2}{32 a^2 \omega^7} + \frac{m^2 \ddot{a}^2}{4 a^2 \omega^5} + \frac{3 m^2 s \ddot{s}}{4 \omega^5} \nonumber \\*
&&- \frac{s \ddot{s}}{8 \omega^3} - \frac{7 m^4 \dot{a} a^{(3)}}{8 a^2 \omega^7} + \frac{15 m^2 \dot{a} a^{(3)}}{32 a^2 \omega^5} + \frac{m^2 a^{(4)}}{16 a \omega^5} \ .\eea


\begin{thebibliography}{99}

\bibitem{parker-toms}L.~Parker and D.~J.~Toms, {\it Quantum Field Theory in Curved Spacetime: Quantized Fields
and Gravity}, Cambridge University Press, Cambridge, England (2009).
\bibitem{Waldbook} R.~M.~Wald, {\it Quantum Field Theory in Curved Space-time and Black Hole Thermodynamics}, University of Chicago Press, Chicago, (1994).
\bibitem{fulling} S.~Fulling, {\it Aspects of Quantum Field Theory in Curved Space-Time}, Cambridge University Press, Cambridge, England (1989).
\bibitem{birrell-davies} N.~D.~Birrell  and P.~C.~W.~Davies, {\it Quantum Fields in Curved Space}, Cambridge University Press, Cambridge, England (1982).
\bibitem{parker66} L.~Parker, {\it The creation of particles in an expanding universe}, Ph.D. thesis, Harvard University (1966). For an historical overview see: L. Parker, {\it J.~Phys.~Conf.~Ser.} {\bf  600} no.1, 012001 (2015);  {\it J.~Phys.~A} {\bf 45}, 374023 (2012).
\bibitem{parker68} L.~Parker, {\it Phys.~Rev.~Lett.} {\bf 21}, 562 (1968); {\it Phys.~Rev.~D} {\bf 183}, 1057 (1969); {\it Phys.~Rev.~D} {\bf 3}, 346 (1971).
\bibitem{parker-fulling} L.~Parker and S.~A.~Fulling, {\it Phys.~Rev.~D} {\bf 9}, 341 (1974); S. A. Fulling and L. Parker, {\it Ann.~Phys.} (N.Y.) {\bf 87}, 176 (1974); S. A. Fulling, L. Parker and B. L. Hu, {\it Phys. Rev. D} {\bf 10}, 3905 (1974). 
\bibitem{Bunch80} T.~S.~Bunch, {\it J.~Phys.~A} {\bf13}, 1297 (1980). 
\bibitem {Anderson-Parker} P.~R.~Anderson and L.~Parker, {\it Phys.~Rev.~D} {\bf 36}, 2963 (1987).  
\bibitem{LNT} A.~Landete, J.~Navarro-Salas and F.~Torrenti, {\it Phys~Rev.~D} {\bf 88}, 061501(R) (2013); {\it Phys.~Rev.~D} {\bf 89}, 044030 (2014).
\bibitem{RNT} A.~del Rio, J.~Navarro-Salas and F.~Torrenti, {\it Phys.~Rev.~D} {\bf 90}, 084017 (2014).
\bibitem{Ghosh:2016} S.~Ghosh, {\it Phys.~Rev.~D} {\bf 91}, 124075 (2015); {\it Phys.~Rev.~D} {\bf 93},   044032 (2016).
\bibitem{christensen76} S.~M.~Christensen, {\it Phys.~Rev.~D} {\bf 14}, 2490 (1976).
\bibitem{Christensen78} S.~M.~Christensen, {\it Phys.~Rev.~D} {\bf 17}, 946 (1978).
\bibitem{Birrell} N.~D.~Birrell, {\it Proc.~R.~Soc. B} {\bf 361}, 513 (1978).
\bibitem{rio1} A.~del Rio and  J.~Navarro-Salas, {\it Phys.~Rev.~D} {\bf 91}, 064031 (2015).
\bibitem{Anderson} P. R. Anderson, {\it Phys.~Rev.~D} {\bf 32}, 1302 (1985); {\it Phys.~Rev.~D} {\bf 33}, 1567 (1986); P.~R.~Anderson and W.~Eaker, {\it Phys. Rev. D} {\bf 61}, 024003 (1999); S.~Habib, C.~Molina-Paris   and E.~Mottola, {\it Phys.~Rev.~D} {\bf 61}, 024010 (1999);
J.~D.~Bates and P.~R.~Anderson,  {\it Phys.~Rev.~D} {\bf 82}, 024018 (2010).
\bibitem{Hu-Parker} B.~L.~Hu  and L.~Parker, {\it Phys.~Lett.~A} {\bf 63}, 217 (1977); {\it Phys.~Rev.~D} {\bf 17}, 933 (1978). 
\bibitem{Anderson-Eaker} P.~R.~Anderson and W.~Eaker,  {\it Phys. Rev. D} {\bf 61}, 024003 (1999).
\bibitem{Molina-Paris-Anderson-Ramsey} C.~Molina-Paris, P.~R.~Anderson and S.~A.~Ramsey, {\it Phys. Rev. D} {\bf 61}, 127501 (2000).


   \bibitem{inflation-r} L.~Parker, arXiv:hep-th/0702216. F.~Finelli, G.~Marozzi, G.~P.~Vacca and G.~Venturi,
  {\it  Phys.\ Rev.\ D} {\bf 76}, 103528 (2007). I.~Agullo, J.~Navarro-Salas, G.~J.~Olmo and  L.~Parker, {\it Phys.~Rev.~Lett.} {\bf 103}, 061301 (2009);
{\it Phys.~Rev.~D} {\bf 81}, 043514, (2010); {\it Phys.~Rev.~D} {\bf 84} 107304,  (2011). R.~Durrer, G.~Marozzi and M.~Rinaldi,
  %``On Adiabatic Renormalization of Inflationary Perturbations,''
  {\it Phys.\ Rev.\ D} {\bf 80}, 065024 (2009). G.~Marozzi, M.~Rinaldi and R.~Durrer,
  {\it  Phys.\ Rev.\ D} {\bf 83}, 105017 (2011). M.~Bastero-Gil, A.~Berera, N.~Mahajan and R.~Rangarajan,
    {\it Phys.\ Rev.\ D} {\bf 87},    087302 (2013). A.~del Rio and J.~Navarro-Salas, {\it Phys.~Rev.~D} {\bf 89} 084037 (2014).~L.~Alinea, T.~Kubota, Y.~Nakanishi and W.~Naylor, {\it JCAP} {\bf 1506} no.06,  019 (2015). A.~L.~Alinea, {\it JCAP} {\bf 1610} no.10,  027 (2016). D.~G.~Wang, Y.~Zhang and J.~W.~Chen, {\it Phys.~Rev.~D} {\bf 94}, 044033 (2016).


\bibitem{Woodard} R.~P.~Woodard, {\it Int.~J.~Mod.~Phys.~D} {\bf 23}, 1430020 (2014).  

\bibitem{Markkanen}  T.~Markkanen and A.~Tranberg,
    JCAP {\bf 1308}, 045 (2013).  T. Markkanen and A. Rajantie, {\it JHEP} {\bf 1701}, 133 (2017).
  
  %\bibitem{inflation-r2} .
\bibitem{KLS} L.~Kofman, A.~Linde and A.~Starobinsky, {\it Phys.~Rev.~Lett.}  {\bf 73}, 3195 (1994); {\it Phys.~Rev.~D} {\bf 56}, 3258 (1997).
\bibitem{Anderson-Molina-Paris-Evanich-Cook} P.~R.~Anderson, C.~Molina-Paris, D.~Evanich and G.~B.~Cook,  {\it Phys.~Rev.~D} {\bf 78}, 083514 (2008).
\bibitem{fermionic-preheating} P.~B.~Greene and L.~Kofman, {\it Phys.~Lett.~B} {\bf 448} 6-12 (1999); {\it  Phys.~Rev.~D} {\bf 62}, 123516  (2000); J.~Baacke, K.~Heitmann and C.~Patzold, {\it  Phys.~Rev.~D} {\bf 58}, 125013 (1998); G.~F.~Giudice, M.~Peloso, A.~Riotto and I.~Tkachev, {\it JHEP} {\bf 9908}, 014 (1999); M.~Peloso and L.~Sorbo, {\it JHEP} {\bf 0005}, 016 (2000); J.~Garcia-Bellido and E.~Ruiz-Morales, {\it Phys.~Lett.~B} {\bf 536}, 193-202 (2002).
\bibitem{figueroa} D.~G.~Figueroa, {\it JHEP} {\bf 1411}, 145 (2014).
\bibitem{enqvist-meriniemi-nurmi} K.~Enqvist, T.~Meriniemi and S.~Nurmi,  {\it JCAP} {\bf 1310}, 057 (2013).
\bibitem{SMHiggsLattice} D.~G.~Figueroa, J.~Garcia-Bellido and F.~Torrenti, {\it  Phys.~Rev.~D} {\bf 92}, 083511 (2015); K.~Enqvist, S.~Nurmi, S.~Rusak and D.~Weir, {\it JCAP} {\bf 1602}, 057 (2016).
\bibitem{higgs-preheating} J.~Garcia-Bellido, D.~G.~Figueroa and J.~Rubio, {\it  Phys.~Rev.~D} {\bf 79}, 063531 (2009); F.~L.~Bezrukov, D.~Gorbunov and M.~Shaposhnikov, {\it JCAP} {\bf 0906}, 029 (2009); D.~G.~Figueroa, and C.~T.~Byrnes, {\it Phys.~Lett.~B} {\bf 767}, 272-277 (2017).
\bibitem{Peskin-Schroeder} M.~E.~Peskin and D.~V.~Schroeder, {\it An Introduction to Quantum Field Theory}, Reading, MA: Addison-Wesley, (1995).
\bibitem{Utiyama-DeWitt} R.~Utiyama and B.~S.~DeWitt, {\it J.~Math.~Phys.} {\bf 3}, 608 (1962). 
\bibitem{Srednicki} M.~Srednicki, {\it Quantum Field Theory}, Cambridge University Press, Cambridge, England (2007).  
\bibitem{bunch-parker} T.~S.~Bunch and L.~Parker, {\it Phys.~Rev.~D} {\bf  20}, 2499 (1979).
\bibitem{bunch81} T.~S.~Bunch, {\it Ann. Phys.} NY, {\bf 131}, 118 (1981).   
\bibitem{Baacke-Patzold} J.~Baacke and C.~Patzold, {\it Phys.~Rev.~D} {\bf 62}, 084008 (2000).
\bibitem{Duff} M.~J.~Duff, {\it Class. Quantum Grav.} {\bf 11}, 1387 (1994).
\bibitem{HHR} S.~W.~Hawking, T.~Hertog and H.~S.~Reall, {\it Phys.~Rev.~D} {\bf 63}, 083504 (2001). 
\bibitem{Wald78} R.~M.~Wald, {\it Phys.~Rev.~D} {\bf 17}, 1477 (1978).

  \end{thebibliography}
\end{document}